\documentclass[prd,11pt,tightenlines,noshowpacs,nofootinbib,amsfonts,amsmath,amssymb,superscriptaddress,onecolumn]{revtex4-1}

\usepackage{amssymb,esvect,amsmath,graphicx,latexsym,amsthm,slashed,eso-pic,hyperref}

\usepackage{epsfig,amsmath,amssymb,verbatim,mathrsfs,hyperref}
\usepackage{xspace}
\usepackage{slashed}
\graphicspath{{./}{Figs/}}

\def\beq{\begin{eqnarray}}
\def\eeq{\end{eqnarray}}
\def\bea{\begin{eqnarray}}
\def\eea{\end{eqnarray}}

\def\gev{\, {\rm GeV}}
\def\mev{\, {\rm MeV}}

\newcommand{\gsim}{\lower.7ex\hbox{$\;\stackrel{\textstyle>}{\sim}\;$}}
\newcommand{\lsim}{\lower.7ex\hbox{$\;\stackrel{\textstyle<}{\sim}\;$}}
\def\mpl{M_{\rm Pl}}

\newcommand{\nnmb}{\nonumber}
\newcommand{\del}{\partial}

\newcommand{\lrf}[2]{\left(\frac{#1}{#2}\right)}

\newcommand{\lag}{\mathscr{L}}

\newcommand{\zpp}{0^{++}}
\newcommand{\opm}{1^{+-}}
\newcommand{\lsvr}{\langle\sigma v\rangle}

\begin{document}
\title{Non-Abelian Dark Forces and the Relic Densities of Dark Glueballs}
\author{Lindsay Forestell}
\affiliation{TRIUMF, 4004 Wesbrook Mall, Vancouver, BC V6T 2A3, Canada}
\affiliation{Department of Physics and Astronomy, University of British Columbia, Vancouver, BC V6T 1Z1, Canada}
\author{David E. Morrissey}
\affiliation{TRIUMF, 4004 Wesbrook Mall, Vancouver, BC V6T 2A3, Canada}
\date{\today}
\author{Kris Sigurdson}
\affiliation{School of Natural Sciences, Institute for Advanced Study, Princeton, NJ 08540, USA}
\affiliation{Department of Physics and Astronomy, University of British Columbia, Vancouver, BC V6T 1Z1, Canada}

\begin{abstract}
Our understanding of the Universe is known to be incomplete, and new gauge forces 
beyond those of the Standard Model might be crucial to describing 
its observed properties.  A minimal and well-motivated possibility is 
a pure Yang-Mills non-Abelian dark gauge force with no direct connection to
the Standard Model.  We determine here the relic abundances of the 
glueball bound states that arise in such theories and 
investigate their cosmological effects.  Glueballs are first formed
in a confining phase transition, and their relic densities are set by
a network of annihilation and transfer reactions.
The lightest glueball has no lighter states to annihilate into, and its
yield is set mainly by $3 \to 2$ number-changing processes which persistently 
release energy into the glueball gas during freeze-out.  The abundances
of the heavier glueballs are dominated by $2\to 2$ transfer reactions,
and tend to be much smaller than the lightest state.  
We also investigate potential connectors between the dark force
and the Standard Model that allow some or all of the dark glueballs to decay.
If the connection is weak, the lightest glueball can be very long-lived or stable
and is a viable dark matter candidate.  For stronger connections, the lightest
glueball will decay quickly, but other heavier glueball states can remain stable
and contribute to the dark matter density.

\end{abstract}

\maketitle


\section{Introduction\label{sec:intro}}

Gauge invariance under the $SU(3)_c\times SU(2)_L\times U(1)_Y$ group 
of the Standard Model~(SM) provides a remarkable description of the 
non-gravitational forces of Nature.  Yet, our knowledge of the Universe 
is incomplete and new gauge forces beyond those of the SM may 
be crucial to describing the laws of physics.
The existence of such forces is highly constrained 
if they couple significantly to SM matter unless they have 
an associated mass scale (such as from confinement or the Higgs mechanism) 
well above a TeV~\cite{Khachatryan:2014fba,Aad:2015ipg}.  
In contrast new \emph{dark} gauge forces, with only feeble connections 
to the SM, can exist at energy scales much less than the TeV scale 
(or even be in a massless phase) 
and still be fully consistent with existing experimental 
bounds~\cite{Pospelov:2008zw,Bjorken:2009mm,Essig:2013lka}.  
Such dark forces may also be related to the cosmological 
dark matter~\cite{Pospelov:2007mp,ArkaniHamed:2008qn,Pospelov:2008jd}.

  Abelian dark forces have been studied in great detail and 
have the novel property that they can connect to the SM at 
the renormalizable level through gauge kinetic 
mixing~\cite{Okun:1982xi,Holdom:1985ag}.
Limits on the existence of such a kinetically-mixed \emph{dark photon}
have been obtained from existing experimental searches and astrophysical
and cosmological observations for a range of dark photon masses spanning 
many orders of magnitude~\cite{Pospelov:2008zw,Bjorken:2009mm,Essig:2013lka}.  
An exciting dedicated experimental program
to search for dark photons is also underway~\cite{Bjorken:2009mm,Essig:2013lka}.

  Non-Abelian dark forces have received somewhat less attention.  
As gauge invariance forbids the simple kinetic-mixing interaction with the SM, 
it is less clear how they might connect to the SM.  
Even so, non-Abelian dark forces are well motivated and arise
in many contexts including string theory constructions~\cite{Blumenhagen:2005mu}, 
in models of dark matter~\cite{Nussinov:1985xr,Barr:1990ca,Faraggi:2000pv,Gudnason:2006ug,Baumgart:2009tn,Kribs:2009fy,Feng:2011ik,Cline:2013zca,Boddy:2014yra,Boddy:2014qxa,Yamanaka:2014pva,Soni:2016gzf}, 
baryogenesis~\cite{An:2009vq,Frandsen:2011kt,Barr:2015lya},
theories of neutral naturalness~\cite{Chacko:2005pe,Burdman:2006tz}, 
and within the \emph{hidden valley} 
paradigm~\cite{Strassler:2006im,Juknevich:2009ji,Juknevich:2009gg}.  
Non-Abelian dark forces can also lead to very different phenomenological
effects compared to their Abelian counterparts owing to the requisite
self-interactions among the corresponding gauge bosons and their potential
for a confining phase transition at low energies.

The minimal realization of a non-Abelian dark force is a pure Yang-Mills
theory with simple gauge group $G_x$.  Such a theory is expected to confine
at the characteristic energy scale $\Lambda_x$, with the elementary dark
gluons binding into a spectrum of colour-neutral dark glueballs of mass 
$m\sim \Lambda_x$~\cite{Jaffe:1975fd}. 
These dark states may have significant cosmological effects even 
when their connection to the SM is too small to be detected 
in laboratory experiments.  For very small values of $\Lambda_x$,
dark gluons can act as self-interacting dark 
radiation~\cite{Jeong:2013eza,Blinov:2014nla,Buen-Abad:2015ova,Reece:2015lch},
and can be consistent with existing constraints provided their effective
temperature is somewhat lower than the SM plasma.
With larger $\Lambda_x$, the glueballs will contribute to the density
of dark matter if they are long-lived~\cite{Faraggi:2000pv,Feng:2011ik,Boddy:2014yra,Boddy:2014qxa,Soni:2016gzf}, or they may lead
to observable astrophysical or cosmological signals if they decay 
at late times~\cite{Faraggi:2000pv,Feng:2011ik,Soni:2016gzf}.

Assessing the cosmological impact of massive dark glueballs requires
a precise knowledge of their relic abundances.  The primary goal of this work 
is to compute these abundances and map out the ranges of parameters 
where one or more dark glueball states might constitute all or some 
of the observed dark matter.  We focus mainly on $G_x=SU(3)$,
but we also comment on how our results can be applied to other
non-Abelian gauge groups.
In a future companion paper we will describe in detail the cosmological effects 
of both stable and unstable primordial glueball populations and use 
them to constrain the existence of non-Abelian dark forces~\cite{us}.

Starting from an early Universe containing a thermal
plasma of dark gluons with temperature $T_x > \Lambda_x$, 
typically different than the temperature of the SM plasma, 
dark glueballs will be formed in a phase transition
as the temperature of the dark sector falls below the confinement scale, 
$T_x \lesssim \Lambda_x$.  Since glueball number is not conserved, 
the number densities of the glueball states will then track their equilibrium 
values so  long as their $2\to 2$ and $n\to 2$ interaction rates are fast
relative to the Hubble expansion rate.  The key difference compared to
standard freeze-out is that without direct annihilation
or rapid decays to SM or lighter hidden states, the overall
chemical equilibrium of the dark glueballs will be maintained primarily by 
$3\leftrightarrow 2$ number-changing 
reactions~\cite{Carlson:1992fn,Hochberg:2014dra,Kuflik:2015isi}.  
Moreover, if the hidden glueballs do not have a kinetic equilibration
with the SM or a bath of relativistic hidden states, the energy released 
by the $3\to 2$ annihilations will cause the remaining
glueballs to cool much more slowly than they would 
otherwise~\cite{Carlson:1992fn}.
Together, these two effects produce freeze-out yields with a much
different dependence on the underlying parameters of the theory
than the typical freeze-out paradigm of annihilation into 
light relativistic particles.

Previous works have studied the effects of $3\to 2$ annihilation 
and self-heating in general massive 
self-coupled sectors~\cite{Carlson:1992fn,Hochberg:2014dra,Kuflik:2015isi}.
The specific application of these processes to dark glueballs has also
been studied in Refs.~\cite{Boddy:2014yra,Boddy:2014qxa,Soni:2016gzf}.  
We expand upon these works in two ways.  
First, we investigate possible effects of the confining
phase transition on the final glueball yields.\footnote{These effects
were studied in a slightly different context in Ref.~\cite{Garcia:2015loa}.} 
And second, we compute the freeze-out abundances of the heavier
glueball states in addition to the lightest mode.  We also show that
when the glueballs are connected to the SM, the heavier relic glueball
states can sometimes have a greater observational effect than the lightest mode.

  Following this introduction, we discuss the general properties of 
dark glueballs in Section~\ref{sec:gb}.  
Next, we study the freeze-out of the lightest glueball 
in Section~\ref{sec:fo} and investigate the effects of 
the confining phase transition.  
In Section~\ref{sec:multi} we extend our freeze-out analysis to include 
the heavier glueball states.  The possibility of dark glueball dark matter 
is studied in Section~\ref{sec:dm}, as well as additional constraints 
that may be placed on general dark forces when a connection to the SM 
is added.  We give brief concluding remarks in Section~\ref{sec:conc}.

\section{Glueball Spectrum and Interactions\label{sec:gb}}

  The spectrum of glueballs in pure $SU(N)$ gauge theories has
been studied extensively using both analytic models and
lattice calculations~\cite{Mathieu:2008me}.  Stable glueballs 
are classified according to their masses and their quantum
numbers under angular momentum~($J$), parity~($P$), 
and charge conjugation~($C$).
The lightest state is found to have 
$J^{PC}=0^{++}$~\cite{Morningstar:1999rf,Meyer:2004gx,Chen:2005mg},
as expected based on general grounds~\cite{West:1995ym}, 
but a number of stable states with other $J^{PC}$ values 
are seen as well.  In this section we summarize
briefly the expected spectrum of glueballs and we estimate
how they interact with each other.

\subsection{Masses}

Much of what is known about the spectrum of glueballs in $SU(N)$ gauge
theories comes from lattice calculations.  It is conventional
to express these masses in terms of a length scale $r_0$ corresponding
to where the gauge potential transitions from Coulombic 
to linear~\cite{Sommer:1993ce,Guagnelli:1998ud}, or in terms
of the confining string tension $\sqrt{\sigma}$.
Both of these quantities can be related to the energy scale 
$\Lambda_{\overline{MS}}$ where the running gauge coupling
becomes strong~\cite{Gockeler:2005rv}.
For $SU(3)$ (with zero flavors), they are given by
$r_0\Lambda_{\overline{MS}} = 0.614(2)(5)$~\cite{Gockeler:2005rv}
and $r_0\sqrt{\sigma} = 1.197(11)$~\cite{Guagnelli:1998ud,Teper:1998kw}.
To facilitate connections with modern lattice calculations,
we will express the glueball masses in terms of $1/r_0$
and define the strong coupling scale as the mass of the
lightest $\zpp$ glueball, $\Lambda_x \equiv m_\zpp$.

Assuming conserved $P$ and $C$ in the dark sector,
the dark glueballs will have definite $J^{PC}$ quantum numbers.
In Table~\ref{tab:gb} we list the spectra of $SU(N)$ glueballs
for $N=2$ and $N=3$ determined in lattice studies in units of $r_0$.  
The $N=3$ glueballs in the table correspond to all the known stable states, 
with the masses listed taken from Ref.~\cite{Morningstar:1999rf}.  
Listings for the $N=2$ case are based on Ref.~\cite{Teper:1998kw},
have significantly larger fractional uncertainties, and may not
give a complete accounting of all the stable states.  Note that
the absence of $C$-odd states is expected for $SU(2)$ 
and other Lie groups with a vanishing 
$d^{abc} = tr(t^a\{t^b,t^c\})$ 
symbol (where $t^a$ is the generator of the 
fundamental representation)~\cite{Jaffe:1985qp,Mathieu:2008me,Juknevich:2009ji}.

\begin{table}[ttt]
\beq
\begin{array}{c|c|c}
J^{PC}&~m\,r_0\:(N=2)~&~m\,r_0\:(N=3)~\\
\hline
0^{++}&4.5(3)&4.21(11)\\
2^{++}&6.7(4)&5.85(2)\\
3^{++}&10.7(8)&8.99(4)\\
0^{-+}&7.8(7)&6.33(7)\\
2^{-+}&9.0(7)&7.55(3)\\
1^{+-}&-&7.18(3)\\
3^{+-}&-&8.66(4)\\
2^{+-}&-&10.10(7)\\
0^{+-}&-&11.57(12)\\
1^{--}&-&9.50(4)\\
2^{--}&-&9.59(4)\\
3^{--}&-&10.06(21)\\
\end{array}
\nnmb
\eeq
\caption{Masses of known stable glueballs in $SU(2)$~\cite{Teper:1998kw}
and $SU(3)$~\cite{Morningstar:1999rf}.
\label{tab:gb}}
\end{table}

  Glueball spectra for $SU(N>3)$ have also been investigated
on the lattice~\cite{Lucini:2004my,Lucini:2010nv}.  The (lowest-lying) 
glueball masses are found to scale with $N$ according to
\beq
r_0\;m(N) \simeq P + Q/N^2 \ ,
\eeq
with $P$ and $Q$ on the order of unity.  These corrections
are found to be numerically modest for $N>3$, and the glueball spectrum
for larger $N$ appears to be similar to $N=3$.  Extrapolations to large
$N$ also find that $r_0\sqrt{\sigma} \simeq 1.2$ remains 
nearly constant~\cite{Teper:1998kw}, while 
the strong-coupling scale decreases smoothly 
to $r_0\Lambda_{\overline{MS}} \simeq 0.45$~\cite{Lucini:2008vi}.
A further variation on $SU(N)$ theories is the addition of a
non-zero topological theta term.  This violates $P$ and $T$ explicitly,
shifts the string tension $\sqrt{\sigma}$ and glueball 
masses~\cite{Vicari:2008jw}, and induces mixing between glueball states 
with different $P$ quantum numbers~\cite{Vicari:2008jw,Gabadadze:2004jq}.

The glueballs for other non-Abelian gauge groups have not been studied
in as much detail on the lattice, but a few specific features are 
expected based on general arguments.  As mentioned above,
there are no $C$-odd states for $SU(2)$, $SO(2N+1)$, or $Sp(2N)$ 
due to their vanishing $d^{abc}$ 
coefficient~\cite{Jaffe:1985qp,Mathieu:2008me,Juknevich:2009ji}.
For $SO(2N)$, $SO(4) \cong SU(2)\times SU(2)$ and $SO(6)\cong SU(4)$
reduce to previous cases, while for $2N > 6$ the $C$-odd states are
expected to be significantly heavier than the lowest $C$-even 
glueballs~\cite{Juknevich:2009ji}.  This follows from the fact that
the minimal gluon operators giving rise to the $C$-odd states 
for the groups have mass dimension $2N$~\cite{Jaffe:1985qp},
and higher-dimension gluon operators are generally expected to lead to 
heavier glueball states~\cite{Jaffe:1985qp,Morningstar:1999rf,Juknevich:2009ji}.

  In this study we concentrate on $SU(N)$ glueballs with $P$ and
$C$ conservation in the dark sector.  However, other non-Abelian gauge groups
could be realized in nature~\cite{Blumenhagen:2005mu}, 
and we comment on these more general scenarios 
when they lead to important phenomenological distinctions.

\subsection{Interactions}

  Dark glueball freeze-out in the early Universe depends on the
cross sections for $2\to 2$ and $3\to 2$ glueball reactions. 
Since glueball self-interactions are expected to be weak
in the limit of very large $N$~\cite{'tHooft:1973jz,Witten:1979kh}, 
it should be possible to calculate these cross sections reliably 
in this limit in perturbation theory using the interactions specified by 
a glueball effective Lagrangian.
These interactions have not been obtained in lattice calculations.
Instead, we estimate them based on large-$N$ 
scaling~\cite{'tHooft:1973jz,Witten:1979kh} 
and na\"ive dimensional analysis~(NDA)~\cite{Manohar:1983md,Cohen:1997rt}

  The most important state for the freeze-out calculation is
the lightest $J^{PC}=0^{++}$ glueball $\phi$.  Using NDA
and large-$N$, we estimate its leading self-interactions to be
\beq
\lag_{eff} &\supset& \frac{1}{2}(\del\phi)^2 - \frac{a_2}{2!}\Lambda_x^2\phi^2
\label{eq:leff}
-\frac{a_3}{3!}\lrf{4\pi}{N}\Lambda_x\,\phi^3
-\frac{a_4}{4!}\lrf{4\pi}{N}^2\phi^4
+ \ldots
\eeq
where the coefficients $a_i$ are expected to be of order unity.
This form matches the NDA scaling of Ref.~\cite{Boddy:2014yra} as well 
as the $1/N$ counting of Ref.~\cite{Soni:2016gzf}

  Applying this form to $2\to 2$ elastic scatterings of the $\zpp$ state
with mass $m_x$, we estimate
\beq
\sigma_{2\to 2}v \simeq \frac{A}{4\pi}\lrf{4\pi}{N}^4\frac{\beta}{s} \ ,
\label{eq:22}
\eeq
where $A$ is dimensionless and close to unity,
$s$ is the square of the center-of-mass energy, and $\beta = \sqrt{1-4m_x^2/s}$.
The same arguments applied to $3\to 2$ processes at low momentum give
\beq
\sigma_{3\to 2}v^2 \simeq \frac{B}{(4\pi)^3}\lrf{4\pi}{N}^6\frac{1}{m_x^5} \ ,
\label{eq:32}
\eeq
with $B$ also close to unity.
These cross sections are at the limit of perturbative unitarity
for small $N$ but become moderate for $N \gtrsim 4\pi$, reflecting
the expected transition to weak coupling in 
this regime~\cite{Witten:1979kh}.
In the analysis to follow we set $A=B=1$, and we generalize 
the cross section estimate for $2\to 2$ interactions to more general processes
involving other glueball states using the same NDA and large-$N$ arguments.

\section{Freeze-out of the Lightest Glueball\label{sec:fo}}

  Having reviewed the properties of glueballs, we turn next to
investigate their freeze-out dynamics in the early Universe.
In this section we study the thermodynamic decoupling of the lightest 
$\zpp$ glueball in a simplified single-state model.  We also
discuss the confining transition in which the glueballs are formed
and investigate how it might modify the glueball relic density.
The freeze-out of heavier glueballs will be studied in the section to follow.

  Throughout our analysis, we assume that the dark glueballs are
thermally decoupled from the SM during the freeze-out process
but maintain a kinetic equilibrium among themselves.  
This implies that the entropy of the dark sector is conserved 
separately from the visible sector, up to a possible increase
during a first-order confining phase transition.  This motivates
the definition
\beq
R \equiv \frac{s_x}{s} = \text{constant} \ ,
\label{eq:entropy}
\eeq  
where $s_x$ is the entropy density of the dark sector \emph{after the
confining transition} and $s$ is that of the visible.  
The value of $R$ is an input to our calculation,
and may be regarded as an initial condition set by the relative reheating
of the dark and visible sectors after inflation if they were never in
thermal contact~\cite{Boddy:2014yra,Adshead:2016xxj}, 
or by the thermal decoupling of the sectors if they once 
were~\cite{Kuflik:2015isi}.  Since inflation can potentially reheat
the dark and visible sectors very asymmetrically, we consider a broad
range of $R \in [10^{-12},10^{-3}]$.

\bigskip

\subsection{Single-State Model}

  Consider first a dark sector consisting of a single real scalar
$\phi_x$ with mass $m_x$, $2\to 2$ and $3\to 2$ self-interaction
cross sections given by Eqs.~(\ref{eq:22},\ref{eq:32}), and no direct
connection to the SM.  We show below that this is often an accurate 
simplified model for the freeze-out of the
lightest $\zpp$ glueball, even when the heavier glueballs are included.  

The freeze-out dynamics of this model
coincide with the general scenario of Ref.~\cite{Carlson:1992fn}.
Chemical equilibrium of the $\phi_x$ scalar is maintained by 
$3\to 2$ transitions.  These transitions also transfer energy
to the remaining $\phi_x$ particles in the non-relativistic plasma
causing them to cool more slowly than they would if there was a relativistic 
bath to absorb the input heat~\cite{Carlson:1992fn,Kuflik:2015isi}.  
Freeze-out occurs when the $3\to 2$ transition rate becomes too slow 
to keep up with the Hubble expansion.
While this happens, kinetic equilibrium is maintained by $2\to 2$
elastic scattering of glueballs, which is parametrically much faster than
the $3\to 2$ processes at dark-sector temperatures below the scalar mass.

Kinetic equilibrium implies that the number density of $\phi_x$ particles
takes the form
\beq
n_x = \int\!\frac{d^3p}{(2\pi)^3}\left[e^{(E_x-\mu_x)/T_x}-1\right]^{-1} \ ,
\label{eq:nx}
\eeq
where $E = \sqrt{\vec{p}^2+m_x^2}$, and $T_x$ and $\mu_x$ refer 
to the temperature and chemical potential of the $\phi_x$ plasma.
This number density evolves in time according 
to~\cite{Carlson:1992fn,Hochberg:2014dra}
\beq
\dot{n}_x + 3Hn_x = - \langle\sigma_{32}v^2\rangle(n_x^3-n_x^2\bar{n}_x)
\label{eq:32fo} \ ,
\eeq
where $H$ is the Hubble rate (sourced by both the visible and dark sectors),
$\bar{n}_x = n_x(\mu_x\!\to\! 0)$ is the number density in the limit of
zero chemical potential, and the thermally-averaged cross section is
\beq
\langle\sigma_{32}v^2\rangle &=& \frac{1}{\bar{n}_x^3}\,
\int\!d\Pi_1d\Pi_2d\Pi_3\;e^{-(E_1+E_2+E_3)/T_x}\,\sigma_{32}v^2 
\label{eq:sig32tt}\\
&\simeq& 
\frac{1}{(4\pi)^3}\lrf{4\pi}{N}^6\frac{1}{m_x^5} \ ,
\nnmb
\eeq
where $d\Pi_i = g_id^3p_i/(2\pi)^32E_i$ 
and we have used Eq.~\eqref{eq:32} in going to the second line.
The dark-sector entropy is
\beq
T_xs_x = \rho_x+p_x-\mu_xn_x \ ,
\eeq
with the energy density $\rho_x$ and pressure $p_x$ determined by
the same distribution function as $n_x$ in Eq.~\eqref{eq:nx}.
Together, Eqs.~(\ref{eq:entropy},\ref{eq:32fo}) provide two equations
for the two unknowns $T_x(t)$ and $\mu_x(t)$ that can be solved
in conjunction with the Friedmann equation for $H(t)$~\cite{Kolb:1990vq}.

  While the results we present below are based on the numerical
evaluation of Eqs.~(\ref{eq:entropy},\ref{eq:32fo}), it is instructive
to derive an approximate solution for the non-relativistic 
freeze-out process~\cite{Carlson:1992fn}.
For $m_x\gg T_x,\,\mu_x$, the dark-sector entropy density is 
\beq
s_x ~\simeq~ \lrf{m_x}{T_x}n_x \ .
\eeq
This relation is maintained with zero chemical potential 
until freeze-out occurs, after which 
the number density just dilutes with the expansion of spacetime.
Matching these limits and applying Eq.~\eqref{eq:entropy}, the freeze-out
yield is 
\beq
Y_x ~=~ \frac{n_x}{s} ~\simeq~ R/x_x^{fo} \ ,
\eeq
where $x_x^{fo} = m_x/T_x^{fo}$ and $T_x^{fo}$ is the dark temperature
at which chemical equilibrium is lost.  
To determine $x_x^{fo}$, we follow Ref.~\cite{Carlson:1992fn}
and identify freeze-out with the point at which the equilibrium $3\to 2$
rate falls below the fractional rate of change of $n_xa^3$,
which gives
\beq
3H \simeq x_x^{fo}\langle\sigma_{32}v^2\rangle\bar{n}_x^2 \ . 
\eeq
Assuming visible radiation dominates the total energy density
during freeze-out, this implies a visible-sector freeze-out temperature of
\beq
T^{fo} \simeq \bar{n}_x\lrf{x_x^{fo}\mpl\langle\sigma_{32}v^2\rangle}{\sqrt{g_*^{fo}\pi^2/10}}^{1/2} \ ,
\eeq
where $\mpl$ is the reduced Planck mass and $g_*^{fo}$ is the number of
effective energy degrees of freedom in the visible sector~\cite{Kolb:1990vq}
at glueball freezeout.
Combining this with the entropy relation of Eq.~\eqref{eq:entropy}
and the explicit form of $\bar{n}_x$ in the non-relativistic regime, we find 
\beq
(x_x^{fo})^{5/2}\,e^{2x_x^{fo}} = \frac{g_{*S}^{fo}}{180\pi}R
\lrf{m_x^4\mpl\langle\sigma_{32}v^2\rangle}{\sqrt{g_*^{fo}\pi^2/10}}^{3/2} \ ,
\eeq
with $g_{*S}^{fo}$ the number of effective entropy degrees of freedom 
in the visible sector~\cite{Kolb:1990vq} at glueball freezeout.
This relation can be solved iteratively for $x_x^{fo}$.

\begin{figure}[ttt]
\centering
\includegraphics[width=0.6\textwidth]{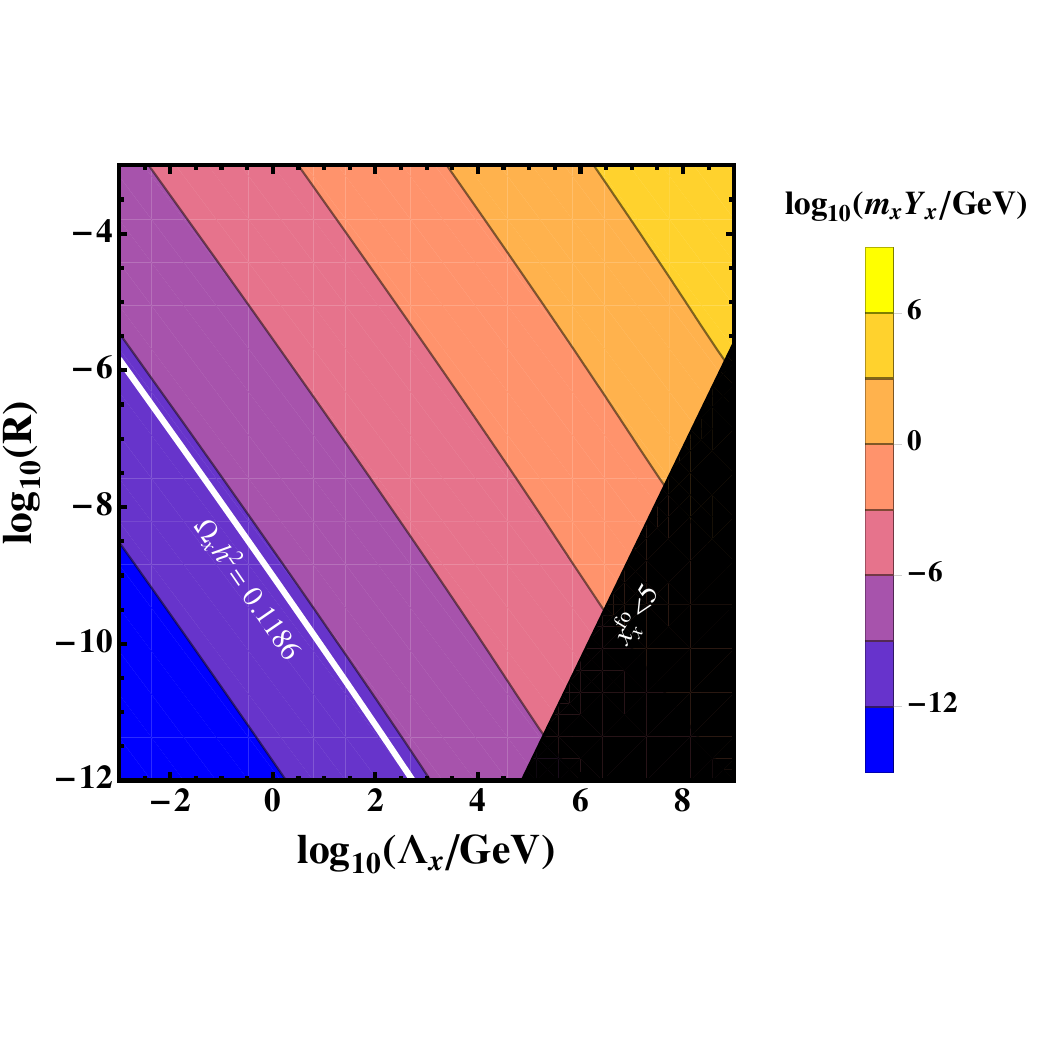}
\vspace{-1.5cm}
\caption{Mass-weighted relic yields in the single-state simplified model
discussed in the text with $N=3$ as a function of the mass $\Lambda_x=m_x$ 
and entropy ratio $R$.  The solid white line indicates where the glueball 
density saturates the observed dark matter abundance 
$\Omega_xh^2 = 0.1186$~\cite{Ade:2015xua}.  The dark masked region
at the lower right indicates where freezeout occurs for $x_x^{fo} < 5$
and our freezeout calculation is not applicable due to the unknown
dynamics of the confining phase transition.
\label{fig:onegb}}
\end{figure}

In Fig.~\ref{fig:onegb} we show the mass-weighted relic yield $m_xY_x$
of $\phi_x$ with $N=3$ as a function of the mass of the lightest glueball 
$\Lambda_x=m_x$ and the dark-to-visible entropy ratio $R$. 
If the lightest glueball is stable, the mass-weighted yield is related
directly to the relic density by
\beq
\Omega_xh^2 = (0.1186)\times\lrf{m_xY_x}{4.322\times 10^{-10}\,\gev} \ .
\eeq
We also indicate on the plot where the relic yield coincides with
the observed dark matter relic density, 
$\Omega_xh^2 = 0.1186$~\cite{Ade:2015xua}. 
The dark shaded region at the lower right corresponds to $x_x^{fo} < 5$.  
As will be discussed below, there is an additional uncertainty 
in the relic abundance in this region when this simplified model 
is applied to dark glueballs, and the present calculation might 
not be applicable here.

\subsection{Dynamics of the Confining Transition}

  Dark glueballs are first formed in the early Universe in a confining
phase transition.  At dark temperatures much larger than the confinement scale, 
$T_x \gg \Lambda_x$, the dark sector can be described as a thermal 
bath of weakly interacting dark gluons with $g_* = 2(N^2-1)$
degrees of freedom.  As $T_x$ cools below $\Lambda_x$
a phase transition occurs with the gluons binding
to form glueballs.  Depending on the nature of the transition and
the interaction rate of the resulting glueballs, this transition
can affect the glueball relic density.

  The nature of the confining transition in pure $SU(N)$ gauge theories has 
been studied in detail on the lattice~\cite{Boyd:1996bx,Beinlich:1996xg,Lucini:2003zr,Lucini:2005vg,Umeda:2008bd,Panero:2009tv,Datta:2010sq,Borsanyi:2012ve,Lucini:2012wq,Francis:2015lha} and in a number of
semi-analytic models (\emph{e.g.} Refs.~\cite{Levai:1997yx,Meisinger:2001cq,Pisarski:2006hz,Gubser:2008yx,Gursoy:2008za,Dumitru:2012fw}).  
The transition is found to be second order for $N=2$, 
weakly first order for $N=3$, and increasingly
first order for $N\geq 4$~\cite{Lucini:2003zr,Lucini:2005vg}.
The dark-sector critical temperature $T_c$ for $N=2\!-\!8$ is fit well by 
the relation~\cite{Lucini:2012wq}
\beq
T_c/\sqrt{\sigma} = 0.5949(17)+0.458(18)/N^2 \ ,
\eeq
where $\sqrt{\sigma} \simeq 1.2/r_0$~\cite{Guagnelli:1998ud}
(or $\sqrt{\sigma} \simeq 2.5\,\Lambda_{\overline{MS}}$~\cite{Gockeler:2005rv,Lucini:2008vi}). 
Note that this is about a factor of five smaller than the mass
of the lightest glueball in Tab.~\ref{tab:gb}.
For $N> 2$ where the transition is found to be first-order, 
the latent heat $L_h$ scales
according to~\cite{Datta:2010sq}
\beq
\frac{L_h}{(N^{2}-1)T_c^4} = 0.388(3) - 1.61(4)/N^2 \ ,
\eeq
while the interface tension between the phases is consistent 
with~\cite{Lucini:2005vg}
\beq
\frac{\sigma_{cd}}{T_c^3} = 0.0138(3)N^2 - 0.104(3) \ .
\eeq
In the confined phase just below the critical temperature,
$0.7\,T_c \lesssim T_x < T_c$, the entropy and pressure are
significantly larger than what is predicted from the known 
glueball states~\cite{Borsanyi:2012ve,Meyer:2009tq}.
Interestingly, this discrepancy can be explained by additional glueball states
with a Hagedorn spectrum corresponding to the excitations of 
a bosonic closed string~\cite{Meyer:2009tq,Buisseret:2011fq,Caselle:2015tza},
in agreement with the model of Ref.~\cite{Isgur:1984bm}.  
The lattice studies of Refs.~\cite{Ishii:2002ww,Meng:2009hh} also suggest that 
the lowest-lying glueball pole masses persist nearly unchanged up
to $T_c$ (although see Ref.~\cite{Caselle:2013qpa} for a different conclusion).

  Much less is known about the non-equilibrium properties of
the $SU(N)$ confining transition such as the nucleation temperature and rate.  
An estimate of the nucleation rate in the early Universe for $SU(3)$, valid in the
limit of small supercooling, is given in Ref.~\cite{Garcia:2015loa}.
For supercooling by an amount $T_x=(1-\delta)T_c$, they find 
a decay per unit volume of
\beq
\Gamma/V \simeq T_x^4e^{-\Delta F_c/T_x} 
\eeq
with
\beq
\frac{\Delta F_c}{T_x} &\simeq& 
\frac{16\pi}{3}\frac{\sigma_{cd}^3}{L_h^2T_c}\delta^{-2} \\
&\simeq&
2.92\times 10^{-4}\delta^{-2}N^2
\frac{\left(1-7.54/N^2\right)^3}{\left(1-4.15/N^2\right)^2} \ ,
\eeq
where $\Delta F_c$ is the difference between the free energies of the two
phases, and in the second line we have generalized the result 
of Ref.~\cite{Garcia:2015loa} to $SU(N\geq 3)$ using the 
central lattice values of $L_h$ and $\sigma_{cd}$ listed above.  
For moderate $N$, this suggests that nucleation occurs at $T_x$
extremely close to $T_c$ (provided $T_x/T \sim R^{1/3}$ is not too small)
with only a very small injection of entropy.  For very large $N$, 
the nucleation rate becomes small and the assumption of small
supercooling made above breaks down.  This suggests that significant 
supercooling can occur at large $N$, although a full non-perturbative 
calculation of the nucleation rate would be needed to verify this.

  To apply these results to the calculation of relic glueball abundances, 
we assume that the phase transition completes with 
$T_x = T_c \simeq m_{\zpp}/5$~\cite{Megevand:2007sv}
and that the mass spectrum of stable glueballs just after the transition
is the same as at $T_x \to 0$.
The simplified model discussed above can then be used with initial
conditions at $x_x = x_{x}^c \equiv m_x/T_c$, which can be specified completely 
in terms of $R=s_x/s$ and $\mu_x(x_{x}^c)$.
If the $3\to 2$ depletion process is fast relative to the Hubble rate
at $x_x=x_x^c$, the initial chemical potential relaxes quickly to zero
and the final relic density is specified completely by the choice of $R$.
However, if full chemical equilibration does not occur at $x_x = x_x^c$,
a range of $\mu(x_x^c)$ values can be consistent with the equilibration rate
relative to Hubble, and there is an additional uncertainty in the final 
glueball relic density for a given value of the entropy ratio $R$.  

\begin{figure}[ttt]
\centering
\includegraphics[width=0.6\textwidth]{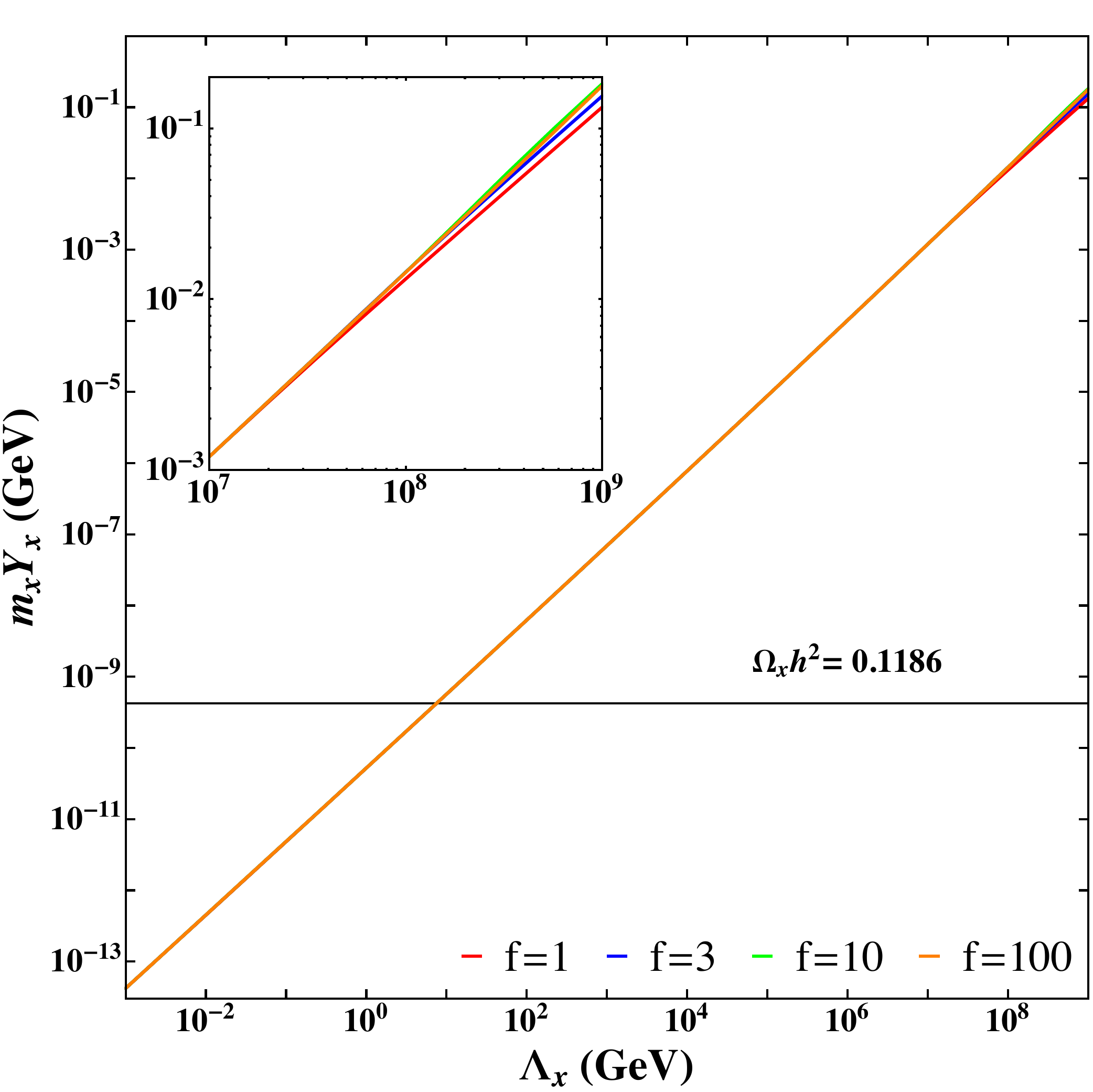}
\caption{Mass-weighted relic yields in the single-state simplified model with the initial density set by $f= Y_x(x_x^c)/Y_x(x_x^c,\mu_x=0)$ at $T_x = \Lambda_x/5$ with $R=10^{-9}$ and $N=3$.  
\label{fig:onegbpt}}
\end{figure}

  To investigate the potential dependence of the relic yield on the 
initial glueball density following the phase transition, we repeat
the freezeout calculation described in the previous section for $G_x=SU(3)$
with different initial glueball densities at $T_x=T_c$ defined by the
ratio $f= Y_x(x_x^c)/Y_x(x_x^c,\mu_x=0)$.  Our results are shown for 
a range of values of $\Lambda_x$  with $R = 10^{-9}$ and $N=3$
in Fig.~\ref{fig:onegbpt}.  For most of the range of $\Lambda_x$ and $R$
of interest, dark freezeout occurs with $x_x^{fo} > x_x^c \simeq 5$ and the final
glueball relic density is insensitive to the initial value after the
phase transition.  Even when $x_x^{fo} < x_x^c$, some residual annihilation
(or creation) typically occurs, and the final density tends to be similar to
$f=1$.  The region in the $\Lambda_x$--$R$ plane in which this additional 
uncertainty is present is indicated by the shaded area in Fig.~\ref{fig:onegb}.

\section{Freeze-Out with Multiple Glueballs\label{sec:multi}}

We turn next to the heavier glueballs above the lightest state.  
Recall from Section~\ref{sec:gb} that multiple stable glueballs
are expected in a confining Yang-Mills theory, with the spectra
found for $SU(2)$ and $SU(3)$ groups listed in Table~\ref{tab:gb}.  
These heavier states lead to new annihilation channels involving the lightest
glueball, and their relic densities can be of cosmological interest.

  The freezeout of the full glueball spectrum involves many states
and a network with numerous reaction channels.
Despite this complexity, we find that the glueball 
relic densities follow a relatively simple pattern with three main features.  
First, the relic density of the lightest glueball is described very well by 
the simplified one-state model presented above provided it freezes out 
while it is significantly non-relativistic.  Second, the relic densities 
of the heavier $C$-even states are typically extremely small relative 
to the lightest glueball.  And third, the total relic density of $C$-odd states
(for $SU(N\geq 3)$ gauge groups) is dominated by the lightest $C$-odd mode 
and is much smaller than the lightest $\zpp$ state but typically larger than 
all the other $C$-even states.  This significant difference arises from the 
conserved $C$ number in the dark sector, which allows coannihilation
of the heavier $C$-even states with the lightest glueball but forbids it
for $C$-odd states.

  In this section we investigate the relic densities of the full set of
glueballs for the dark gauge group $SU(3)$.  We begin by determining
which $2\to 2$ glueball reactions are allowed by $J^{PC}$ conservation
in the dark sector, and we estimate their rates.  Next, we study a simplified
reaction network of $C$-even states that we argue captures the most important
features of the full dynamics.  Finally, we perform a similar analysis
for the $C$-odd states.

\subsection{Glueball Reactions}

\begin{table}[ttt]
\beq
\begin{array}{c|c|c}
~~~~i~~~~&~~~J^{PC}~~~&~~~m_i/m_{\zpp}~\\
\hline\hline
1&\zpp&1.00\\
2&2^{++}&1.39\\
3&3^{++}&2.13\\
\hline
4&0^{-+}&1.50\\
5&2^{-+}&1.79\\
\hline
6&1^{+-}&1.70\\
7&3^{+-}&2.05\\
8&2^{+-}&2.40\\
9&0^{+-}&2.74\\
\hline
10&1^{--}&2.23\\
11&2^{--}&2.27\\
12&3^{--}&2.39
\end{array}\nnmb
\eeq
\caption{List of stable glueball states and mass ratios for $SU(3)$, 
from Ref.~\cite{Morningstar:1999rf}.}
\label{tab:states}
\end{table}

To discuss glueball reactions for $G_x = SU(3)$, it will be convenient
to label the modes in the spectrum by $i=1,2,\ldots,12$
as in Table~\ref{tab:states}.  This table also lists their relative
masses and $J^{PC}$ quantum numbers.  

The specific interactions between glueballs are not known, but all possible
processes consistent with dark-sector $J$, $P$, and $C$ conservation are
expected to be present.  For a $2\to 2$ glueball reaction of the form
$i+j\to k+l$, conservation of $C$ requires
\beq
C_jC_j = C_kC_l \ .
\eeq
This is trivial to apply and rules out a number of reactions.
Conservation of $P$ implies
\beq
P_iP_j = (-1)^LP_kP_l \ ,
\label{eq:parity}
\eeq
where $L$ is the relative orbital angular momentum of the reaction channel.
When identical particles are present, they must also be symmetrized.
In general, it can be shown that there always exists a value of $L$ such that both parity and total angular momentum are conserved unless either 
$J_i=J_j=0$ or $J_k=J_l=0$.  

If the process $i+j\leftrightarrow k+l$ is allowed, it contributes
to the collision term in the Boltzmann equation for glueball $i$ according to
\beq
\Delta\dot{n}_i ~=~ -\langle\sigma v\rangle_{ijkl}n_in_j 
+ \langle\sigma v\rangle_{klij}n_kn_l \ ,
\eeq
where $\langle\sigma v\rangle_{ijkl}$ is the thermally-averaged cross section
and $n_i$ refers to the number density of the $i$-th species.
Assuming kinetic equilibrium is maintained among the glueballs, we have
\beq
n_i &=& g_i\,e^{\mu_i/T_x}(4\pi)m_i^2T_x\,K_2(m_i/T_x) \\
&\simeq& g_i\lrf{m_iT_x}{2\pi}^{3/2}e^{-(m_i-\mu_i)/T_x} \ ,
\eeq
where $T_x$ is the temperature of the glueball bath and $g_i$, $m_i$, and $\mu_i$
are the number of degrees of freedom, mass, and chemical potential of the
type-$i$ glueball.
The thermally-averaged cross-section is given by
\beq
\langle \sigma v\rangle_{ijkl} &=& 
\frac{1}{n_in_j}\,\int\!\frac{d^3p_i}{(2\pi)^3}\int\!\frac{d^3p_j}{(2\pi)^3}\;
g_i\,e^{(\mu_i-E_i)/T_x}g_j\;e^{(\mu_j-E_j)/T_x}(\sigma v)_{ijkl}\\
&=&
\frac{g_ig_j}{\bar{n}_i\bar{n}_j}\,
\int\!\frac{d^3p_i}{(2\pi)^3}\int\!\frac{d^3p_j}{(2\pi)^3}\;
e^{-(E_i+E_j)/T_x}\,(\sigma v)_{ijkl} \ ,
\eeq
where $E_i = \sqrt{m_i^2+\vec{p}_i^2}$ 
and $\bar{n}_i = n_i(\mu_i=0)$.  
Note that the chemical potentials cancel in this expression.  

The reaction $i+j\to k+l$ is either exothermic ($m_i+m_j \geq m_k+m_l$) 
or endothermic ($m_i+m_j < m_k+m_l$).
Equilibration of this process implies $\mu_i+\mu_j = \mu_k+\mu_l$.
Combined with detailed balance, we must have
\beq
\langle \sigma v\rangle_{ijkl}\;\bar{n}_i\bar{n}_j
= \langle \sigma v\rangle_{klij}\;\bar{n}_k\bar{n}_l \ .
\label{eq:exend}
\eeq
Using these relations, the thermally-averaged rates of endothermic
reactions can be estimated based on those of exothermic reactions.

Thermal averaging of cross sections was studied in detail 
in Refs.~\cite{Gondolo:1990dk,Edsjo:1997bg}. 
Generalizing their results slightly and using the large-$N$ and NDA
estimates of interaction strengths, we estimate the thermally-averaged
cross section of an exothermic process $i+j\to k+l$ that proceeds 
at lowest orbital angular momentum level $L$ by
\beq
\langle \sigma v\rangle_{ijkl} ~\simeq~ 
\frac{(4\pi)^3}{N^4}\,\frac{\beta_{ijkl}}{s_{ij}}\,\;
c_L\!\lrf{2}{x_i\!+\!x_j}^{L} \ ,
\label{eq:tavg}
\eeq
where $x_i = m_i/T$,
\beq
s_{ij} = \left(1+\frac{3}{x_{i}+x_j}\right)(m_i+m_j)^2 \ ,
\eeq
along with
\beq
\beta_{ijkl} &=& \frac{2p'_{kl}}{\sqrt{s_{ij}}}
\label{eq:beta}\\
&=& \frac{1}{s_{ij}}\left(s_{ij}^2+m_k^4+m_l^4-2s_{ij}m_k^2-2s_{ij}m_l^2
-2m_l^2m_k^2\right)^{1/2} \nnmb \ ,
\eeq
and the coefficients $c_L$ are~\cite{Gondolo:1990dk}
\beq
c_0 = 1,~~c_1 = 3/2,~~c_2 = 15/8,~~c_3 = 35/16,~~c_4 = 315/128 \ .
\eeq
The first factor in Eq.~\eqref{eq:tavg} contains the couplings,
the second factor describes the kinematics near threshold 
in the non-relativistic limit, while the third is the velocity 
suppression for a process that goes at the $L$-th partial wave.

  The cross-section estimates of Eq.~\eqref{eq:tavg} can be used
to judge which reactions are most significant during freezeout.
The relative effect of the process $i+j\to k+l$ 
(with $j,\,k,\,l \neq i$) on the number density of glueball species $i$ is 
\beq
\frac{|\Delta\dot{n}_i|}{n_i} = \langle\sigma v\rangle_{ijkl}\,n_j \ .
\eeq
In general, this reaction is cosmologically active for 
$|\Delta \dot{n}_i|/n_i > H$.  Scanning over all possible $2\to 2$ reactions 
of $SU(N=3)$ glueballs, we find that in full equilibrium with $x_x > 5$ 
and for every glueball species $i> 1$ there exist multiple number-changing 
$2\to 2$ reactions down to lighter states with $|\Delta\dot{n}_i|/n_i$ 
significantly larger than the corresponding quantity for $3\to 2$ 
annihilation of the lightest glueball.
This implies that \emph{relative chemical equilibrium} is maintained among the
glueballs during and for some time after $3\to 2$ freeze-out, with
\beq
\frac{n_i}{n_j} = \frac{\bar{n}_i}{\bar{n}_j} \simeq 
\frac{g_i}{g_j}\lrf{m_i}{m_j}^{3/2}e^{-(m_i-m_j)/T_x} \ .
\label{eq:crel}
\eeq
Equivalently, the number densities of all species immediately after $3\to 2$
freeze-out are given by their equilibrium values with a common chemical potential.

  Relative chemical equilibrium after $3\to 2$ freezeout
implies further that the relative importance of different $2\to 2$ reactions 
on the subsequent freezeout of the heavier glueballs can be estimated
using their equilibrium number densities.  This allows us to greatly simplify
the set of reaction networks by keeping only the dominant processes
and concentrating exclusively on a few key states.  It turns out to be consistent
and convenient to study the $C$-even and $C$-odd states independently,
and this is the approach we take below.

\subsection{Relic Densities of $C$-Even States}

  The lightest $C$-even glueballs above the lowest mode have
$J^{PC} = 2^{++},\,0^{-+},\,2^{-+}$ and correspond to $i=2,\,4,\,5,$ 
in our labelling scheme.
Scanning over all possible reactions for these states and estimating
their relative effects on the number densities as above, the dominant
interactions near relative equilibrium are found
to form a minimal closed system.  The reaction network for the system
is described by
\beq
\dot{n}_1 + 3H\,n_1 &=& - \langle\sigma_{32}v^2\rangle n_1^2(n_1-\bar{n}_1)
\label{eq:n1b}\\
&&- \frac{1}{2}\langle\sigma v\rangle_{2111}
\left[\lrf{\bar{n}_2}{\bar{n}_1}n_1n_2- n_2^2\right]
\nnmb\\
&&-\langle\sigma v\rangle_{2211}
\left[\lrf{\bar{n}_2}{\bar{n}_1}^2n_1^2-n_2^2\right]
\nnmb\\
&&-\frac{1}{2}\langle\sigma v\rangle_{2214}
\left[\lrf{\bar{n}_2^2}{\bar{n}_1\bar{n}_4}n_1n_4-n_2^2\right]
\nnmb\\
\phantom{I}\nnmb\\
&&-\frac{1}{2}\langle \sigma v \rangle_{2415} \left[\lrf{\bar{n}_2\bar{n}_4}{\bar{n}_1\bar{n}_5} n_1 n_5 - n_2 n_4 \right]
\nnmb \\
\dot{n}_2 + 3H\,n_2 &=&+ \frac{1}{2}\langle\sigma v\rangle_{2111}\left[\lrf{\bar{n}_2}{\bar{n}_1}n_1n_2-n_2^2 \right]
\label{eq:n2b}\\
&&+\langle\sigma v\rangle_{2211}
\left[\lrf{\bar{n}_2}{\bar{n}_1}^2n_1^2-n_2^2\right]
\nnmb\\
&&+\langle\sigma v\rangle_{2214}
\left[\lrf{\bar{n}_2^2}{\bar{n}_1\bar{n}_4}n_1n_4-n_2^2\right]
\nnmb\\
\phantom{I}\nnmb\\
&&+\frac{1}{2}\langle \sigma v \rangle_{2415} \left[\lrf{\bar{n}_2\bar{n}_4}{\bar{n}_1\bar{n}_5} n_1 n_5 - n_2 n_4 \right]
\nnmb \\
&&-\frac{1}{2}\langle \sigma v \rangle_{1512} \left[\lrf{\bar{n}_1\bar{n}_5}{\bar{n}_1\bar{n}_2} n_1 n_2 - n_1 n_5 \right]
\nnmb \\
\dot{n}_4 + 3H\,n_4 &=&- \frac{1}{2}\langle\sigma v\rangle_{2214}
\left[\lrf{\bar{n}_2^2}{\bar{n}_1\bar{n}_4}n_1n_4-n_2^2\right] \ 
\label{eq:n4b}\\
&&+\frac{1}{2}\langle \sigma v \rangle_{2415} \left[\lrf{\bar{n}_2\bar{n}_4}{\bar{n}_1\bar{n}_5} n_1 n_5 - n_2 n_4 \right]
\nnmb \\
\dot{n}_5 + 3H\,n_5 &=&-\frac{1}{2}\langle \sigma v \rangle_{2415} \left[\lrf{\bar{n}_2\bar{n}_4}{\bar{n}_1\bar{n}_5} n_1 n_5 - n_2 n_4 \right]
\label{eq:n5b}\\
&&+\frac{1}{2}\langle \sigma v \rangle_{1512} \left[\lrf{\bar{n}_1\bar{n}_5}{\bar{n}_1\bar{n}_2} n_1 n_2 - n_1 n_5 \right]
\nnmb
\eeq
The factors of $1/2$ appearing here are symmetry factors
for initial states that are not included the standard definition of 
the thermally-averaged cross section~\cite{Edsjo:1997bg}.
They ensure that the summed number density $n_1+n_2+n_4+n_5$ 
is conserved in the absence of $3\to 2$ reactions. 
In addition to these evolution equations, the ratio of entropies
$R=s_x/s$ is conserved after the confining transition at 
$T_x^c \simeq m_x/5$, with the dark sector entropy now extended to 
include all (relevant) glueball modes.

\begin{figure}[ttt]
	\centering
	\includegraphics[width=0.47\textwidth]{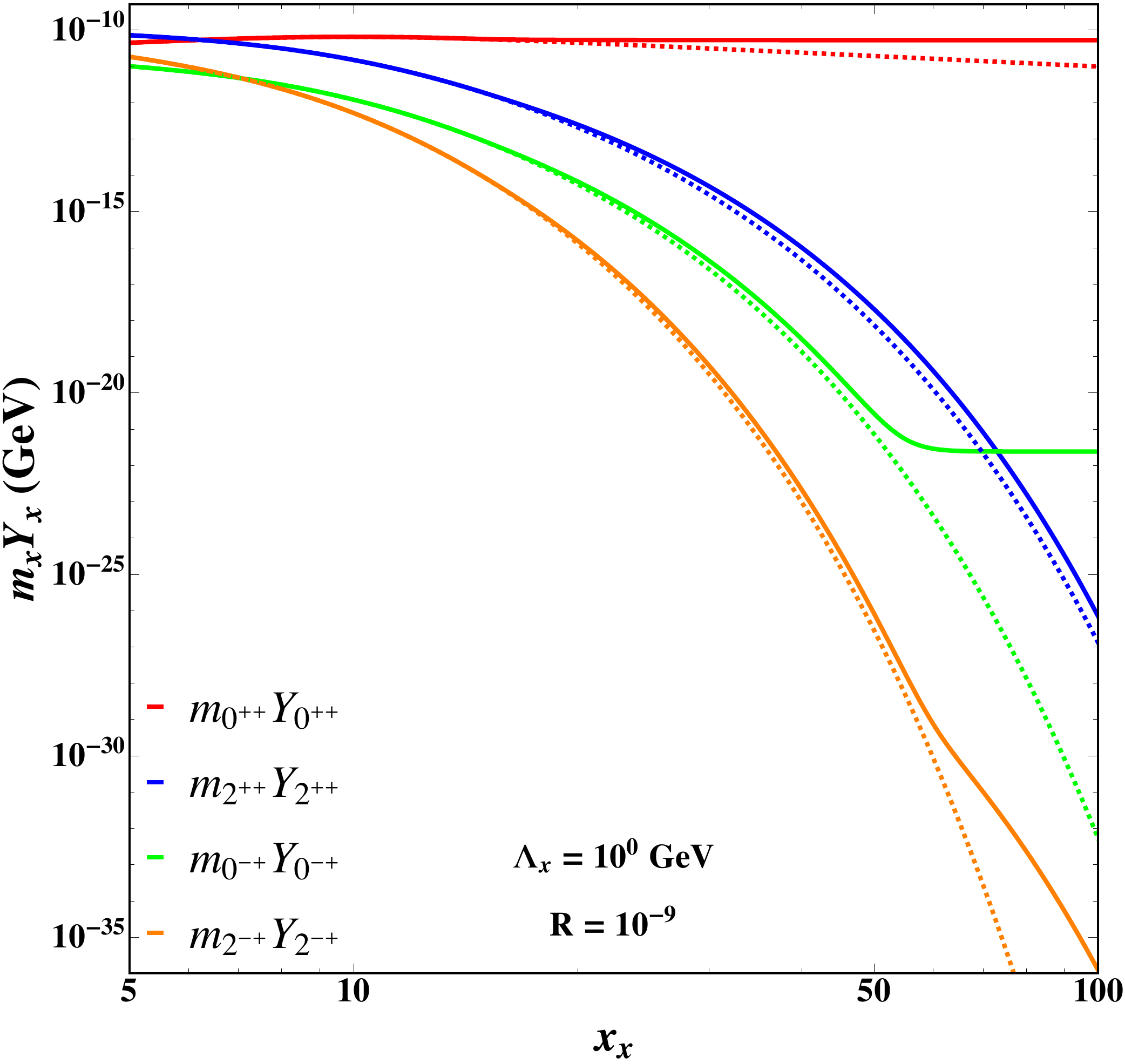}
	\hspace{0.5cm}
	\includegraphics[width=0.47\textwidth]{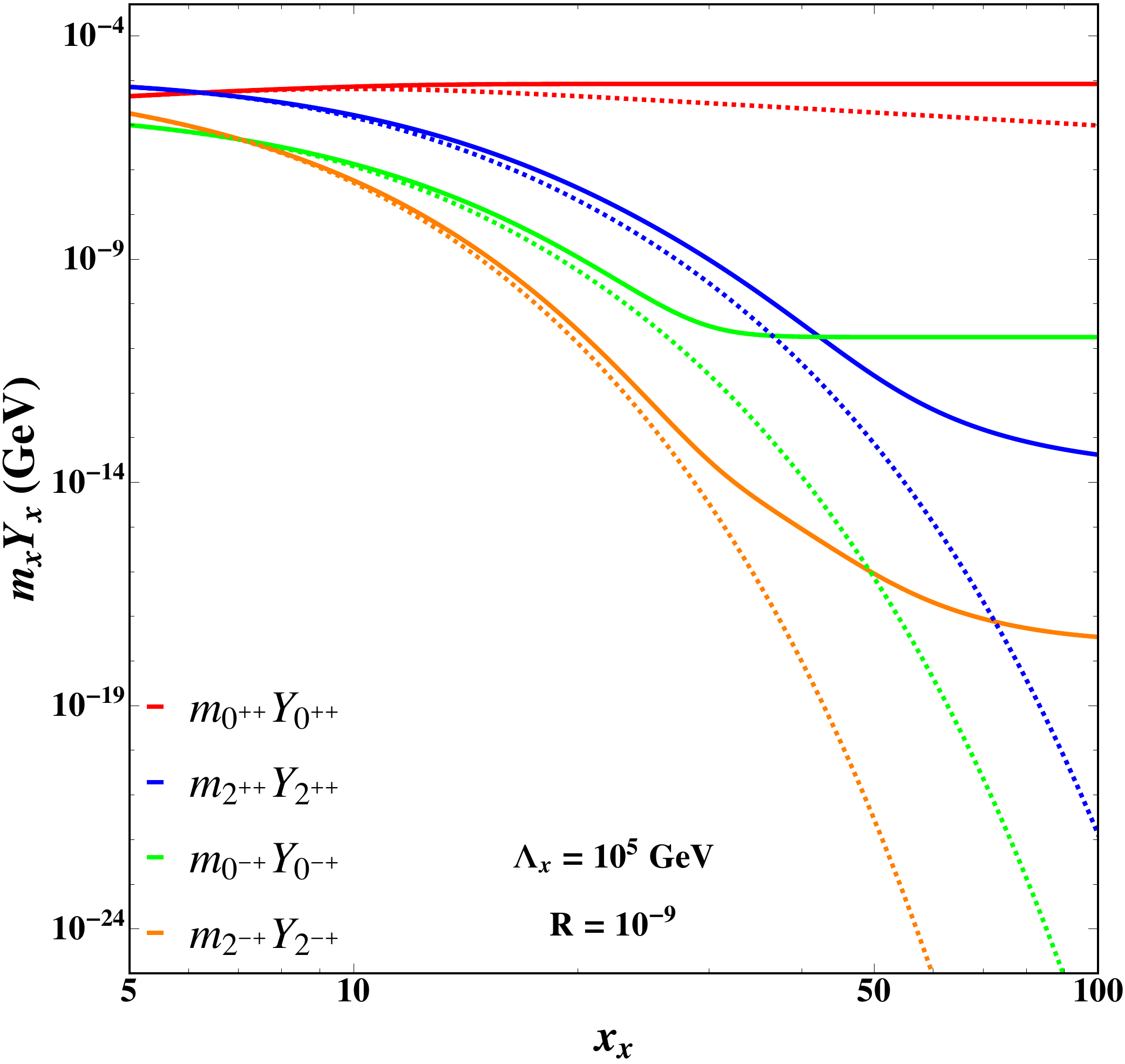}
	\hspace{0.5cm}
	\includegraphics[width=0.47\textwidth]{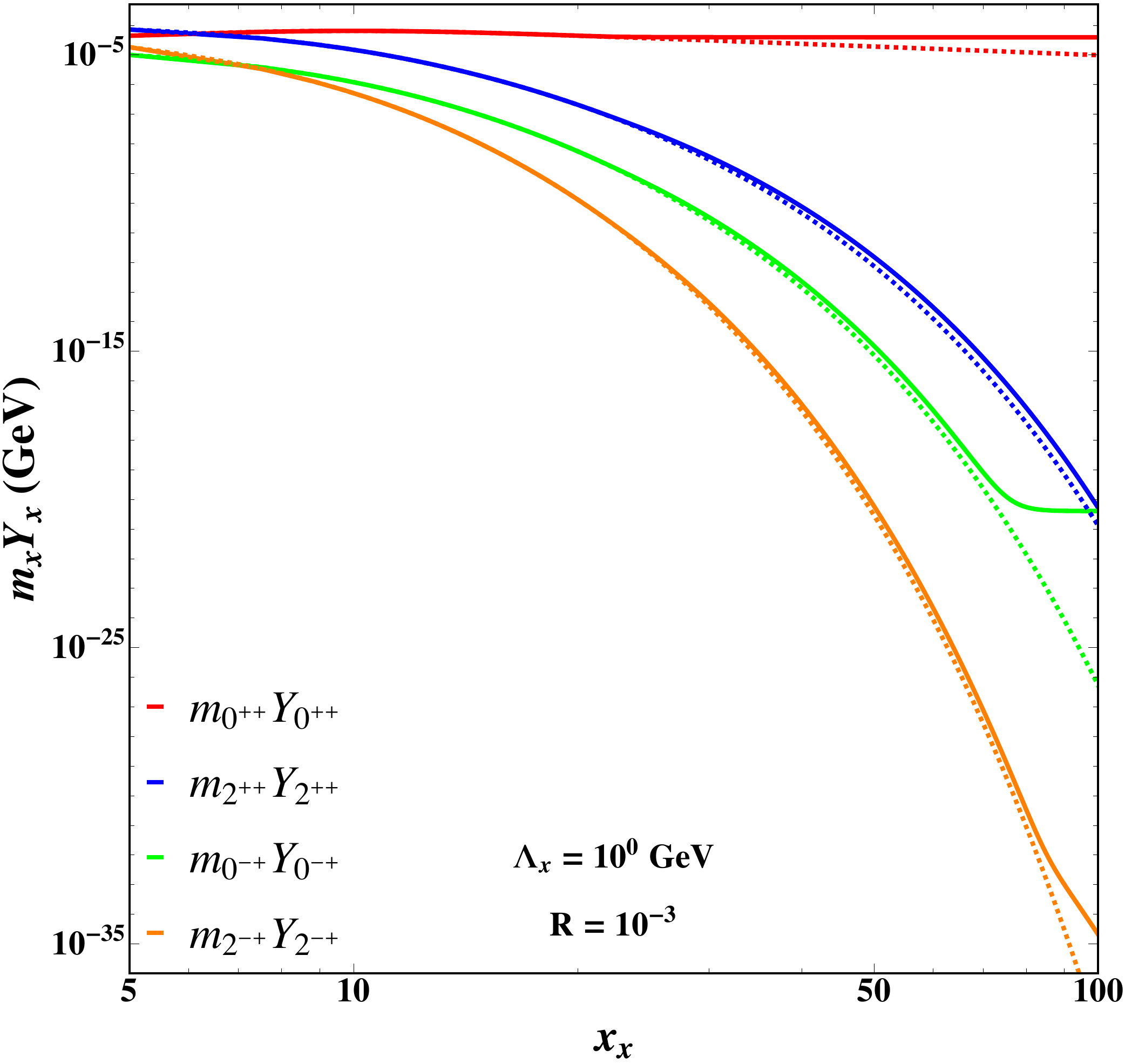}
	\hspace{0.5cm}
	\includegraphics[width=0.47\textwidth]{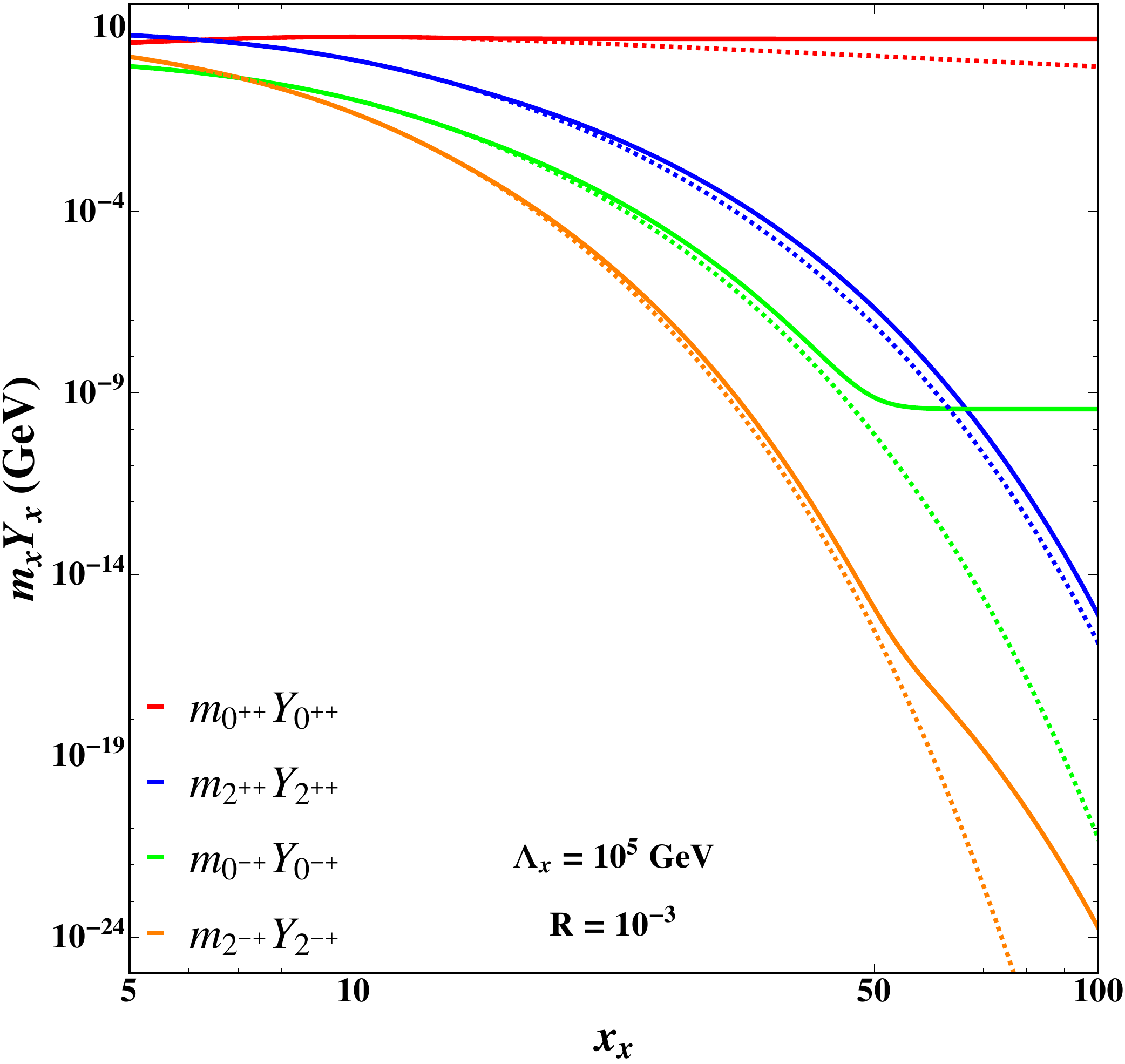}
	\hspace{0.5cm}
	\caption{Mass-weighted relic yields of the four lightest $C$-even glueballs in $SU(3)$, $J^{PC} = 0^{++},\,2^{++},\,0^{-+},\,2^{-+}$, as a function of the dark glueball temperature variable $x_x = m_x/T_x$ computed using the simplified reaction network discussed in the text.  The solid lines show the yields derived from the reaction network while the dashed lines indicate the yields expected if the states were to continue following equilibrium with $\mu_i=0$.  Top left: $(\Lambda_x/\gev,\,R) = (1,\,10^{-9})$. Top right: $(\Lambda_x/\gev,\,R) = (10^5,\,10^{-9})$. Bottom left: $(\Lambda_x/\gev,\,R) = (1,\,10^{-3})$. Bottom right: $(\Lambda_x/\gev,\,R) = (10^5,\,10^{-3})$.
\label{fig:fourgbxY}
}
\end{figure}

\begin{figure}[ttt]
	\centering
	\includegraphics[width=0.47\textwidth]{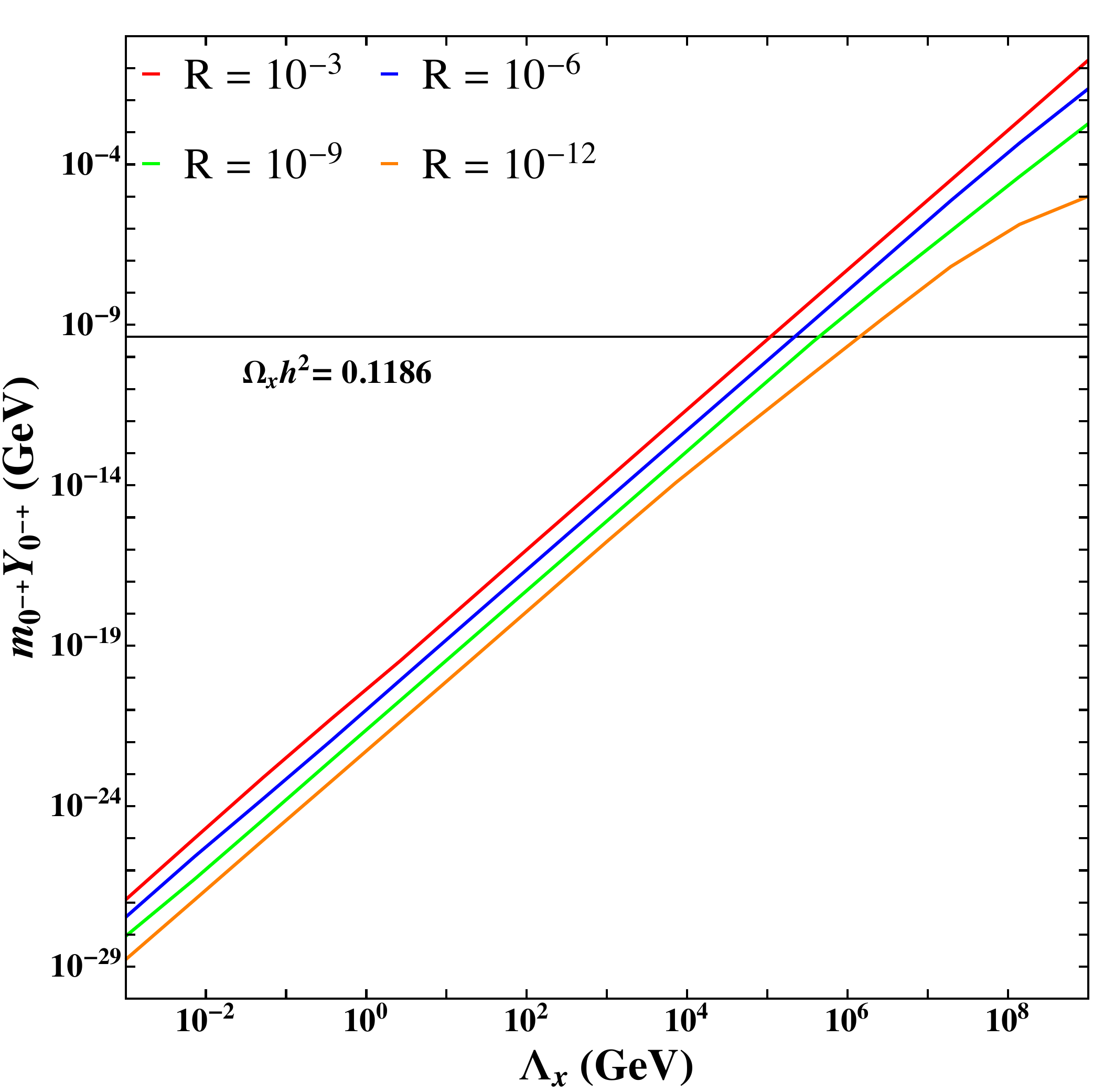}
	\hspace{0.5cm}
	\includegraphics[width=0.47\textwidth]{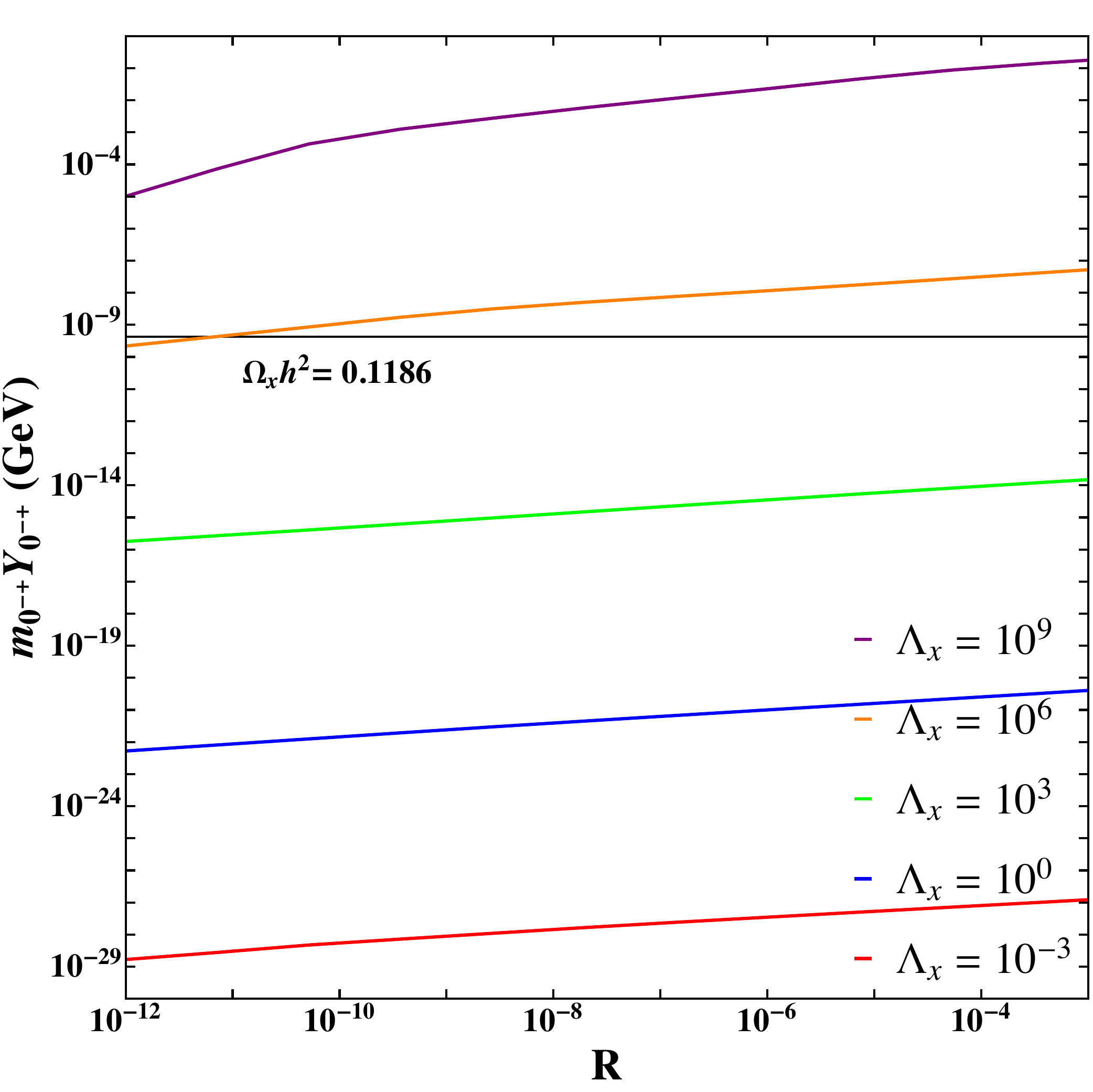}
	\hspace{0.5cm}
	\caption{Mass-weighted relic yields of the $0^{-+}$ dark glueball in $SU(3)$ as functions of $\Lambda_x = m_x$ and $R$, computed using the simplified $C$-even reaction network discussed in the text.  For reference, we also indicate the yield corresponding to the observed dark matter density.  Note that the yield of the $0^{++}$ state is much larger.
\label{fig:fourgb}}
\end{figure}

  Numerical solutions of this system of equations for $SU(3)$ dark glueballs
are shown in Fig.~\ref{fig:fourgbxY} for the parameter values
$(\Lambda_x/\gev,\,R) = (1,\,10^{-9}),\,(10^5,\,10^{-9}),\,(1,\,10^{-3}),\,(10^5,\,10^{-3})$.  In each panel, the evolution of the mass-weighted yields $m_iY_x$
with $x_x=m_x/T_x$ are given by the solid lines, while the dashed lines show
the mass-weighted yield of each species with $\mu_i=0$.  In all four panels,
the lightest $\zpp$ mode is seen to dominate the total glueball relic abundance 
for $x_x \gtrsim 10$.  This abundance is found to match closely with
the value determined by the one-glueball simplified model discussed above.
The much smaller relic abundances of the heavier glueball modes is due to
the efficient coannihilation reactions they experience.  Since these $2\to 2$ 
processes are parametrically faster than the $3\to 2$ annihilations setting 
the $\zpp$ density, relative chemical equilibrium is maintained to 
large values of $x_x$.  This implies a strong exponential suppression
of the heavier glueball densities as in Eq.~\eqref{eq:crel}.

Let us also point out that the $0^{-+}$ state freezes out (of relative
chemical equilibrium) well before the $2^{++}$ and $2^{-+}$ modes, even
though it is heavier than the $2^{++}$.  This can be understood by
examining the relative rates of the depletion reactions for the $0^{-+}$ state;
for $x_x \gtrsim 20$ it is found to be $0^{-+}+\zpp \to 2^{++}+2^{++}$.
Comparing masses, this reaction is found to be endothermic and thus it 
receives an additional rate suppression as discussed 
in Ref.~\cite{D'Agnolo:2015koa}.  The dependence of the $0^{-+}$ ($i=4$)
glueball relic density on $\Lambda_x = m_x$ and $R$ is also shown in 
Fig.~\ref{fig:fourgb}.

\subsection{Relic Densities of $C$-Odd States}

  Freezeout of the $C$-odd glueballs is qualitatively different from that
of the $C$-even modes due to the conservation of $C$ number in the dark sector.
This forbids coannihilation reactions of the $C$-odd states with the relatively
abundant lightest $\zpp$ glueball into final states with only $C$-even modes,
and can lead to a significant relic density for the lightest
$C$-odd $1^{+-}$ state.  

  To see how this comes about, let us split up the labels of the state 
indices defined in Table~\ref{tab:states} according to
\beq
i,j = 1,2,3,4,5 = C\text{-even} 
\ ,~~~~~
a,b = 6,7,\ldots 12 = C\text{-odd} \ ,
\eeq
and let us also define the net $C$-odd density by
\beq
n_{-} = \sum_{a=6}^{12}n_a \ .
\eeq
The net collision term in the Boltzmann equation for $n_-$ is
\beq
\Delta\dot{n}_{-} &=& \sum_a\Delta\dot{n}_a\\
&=& -\sum_{ab,ij}\langle\sigma v\rangle_{abij}\,n_an_b
+ \sum_{ij,ab}\langle\sigma v\rangle_{ijab}\,n_in_j  \ .
\label{eq:cm}
\eeq
The key feature of this expression is that all processes
contributing to the rate of change of $\dot{n}_-$
have two $C$-odd particles either in the initial or the final 
state~\cite{Edsjo:1997bg}.
Using detailed balance, we can rewrite Eq.~\eqref{eq:cm} in the form
\beq
\Delta\dot{n}_- &=& -\lsvr_{6611}n_-^2\left[
\sum_{abij}\left(\frac{\lsvr_{abij}}{\lsvr_{6611}}\frac{n_an_b}{n_-^2}
\,\Theta_+
+ \frac{\lsvr_{ijab}}{\lsvr_{6611}}\frac{n_an_b}{n_-^2}
\frac{\bar{n}_i\bar{n}_j}{\bar{n}_a\bar{n}_b}
\,\Theta_-\right)\right]
\label{eq:bcodd}
\\
&&+ \lsvr_{6611}n_1^2\lrf{\bar{n}_-}{\bar{n}_1}^2\!\left[
\sum_{abij}\left(
\frac{\lsvr_{abij}}{\lsvr_{6611}}
\frac{\bar{n}_a\bar{n}_b}{n_-^2}\frac{n_in_j}{\bar{n}_i\bar{n}_j}
\frac{\bar{n}_1^2}{n_1^2}
\,\Theta_+
+\frac{\lsvr_{ijab}}{\lsvr_{6611}}\frac{n_in_j}{n_-^2}\frac{\bar{n}_1^2}{n_1^2}
\,\Theta_-
\right)\right]
\nnmb \ ,
\eeq
where $\Theta_+ = \Theta(m_a+m_b-m_i-m_j)$ and 
$\Theta_- = \Theta(m_i+m_j-m_a-m_b)$ are step functions to select out exothermic
reactions as appropriate.

Consider the relative sizes of the individual terms in Eq.~\eqref{eq:bcodd}
when relative equilibrium is maintained.  In the first line, 
the first term is on the order of unity for $a=b=6$ but has 
an exponential suppression otherwise
from the factor of $n_an_b/n_-^2$, 
while the second term has an additional exponential 
suppression from the factor $\bar{n}_i\bar{n}_j/\bar{n}_a\bar{n}_b$ 
($m_i+m_j > m_a+m_b$).  Similar arguments also apply to the terms in the second 
line of Eq.~\eqref{eq:bcodd}, noting that 
$\bar{n}_i\bar{n}_jn_1^2 = n_in_j\bar{n}_1^2$ in relative equilibrium,
and only the $a=b=6$ portion of the first term avoids an exponential
suppression.  Indeed, a numerical evaluation of these contributions, 
assuming relative equilibrium and moderate $x_x \gtrsim 10$, confirms
that the $a=b=6$ terms of the $\Theta_+$ pieces dominate the collision term.

  The total $C$-odd density begins to deviate appreciably from the
relative equilibrium value for $\lsvr_{6611}\bar{n}_-^2(n_1/\bar{n}_1)^2 \sim H$.
This occurs well before the $C$-even states freeze out, and also well before
$C$-odd transfer reactions, such as $7+1 \leftrightarrow 2+6$, turn off.
The latter result implies that the relative densities of $C$-odd states
are maintained among themselves (but not the $C$-even states) even
after the net $C$-odd density has frozen out.  Therefore we also
expect $n_6/n_-\to 1$ and $n_{a>6}/n_-\to 0$ provided these processes
turn off at moderate $x_x \gtrsim 10$.  

The net result of this analysis is that it is generally a good
approximation to compute the freezeout of the $C$-odd density using 
a simplified two-state system consisting only of the 
$i=1,\,6$ ($\zpp$ and $1^{+-}$) glueballs.  
Correspondingly, the system of Boltzmann equations is
\beq
\dot{n}_1 + 3Hn_1 &=& -\langle\sigma_{32}v^2\rangle n_1^2(n_1-\bar{n}_1)\\
&& + \lsvr_{6611}\left[n_6^2 - \lrf{{n}_1}{\bar{n}_1}^2\bar{n}_6^2\right]
\nnmb\\
\nnmb\\
\dot{n}_6 + 3Hn_6 &=& - \lsvr_{6611}\left[n_6^2 - \lrf{{n}_1}{\bar{n}_1}^2\bar{n}_6^2\right] \ .
\eeq
Corrections to this estimate are expected to be of order unity,
which is well within the uncertainties on the cross sections.

\begin{figure}[ttt]
	\centering
	\includegraphics[width=0.47\textwidth]{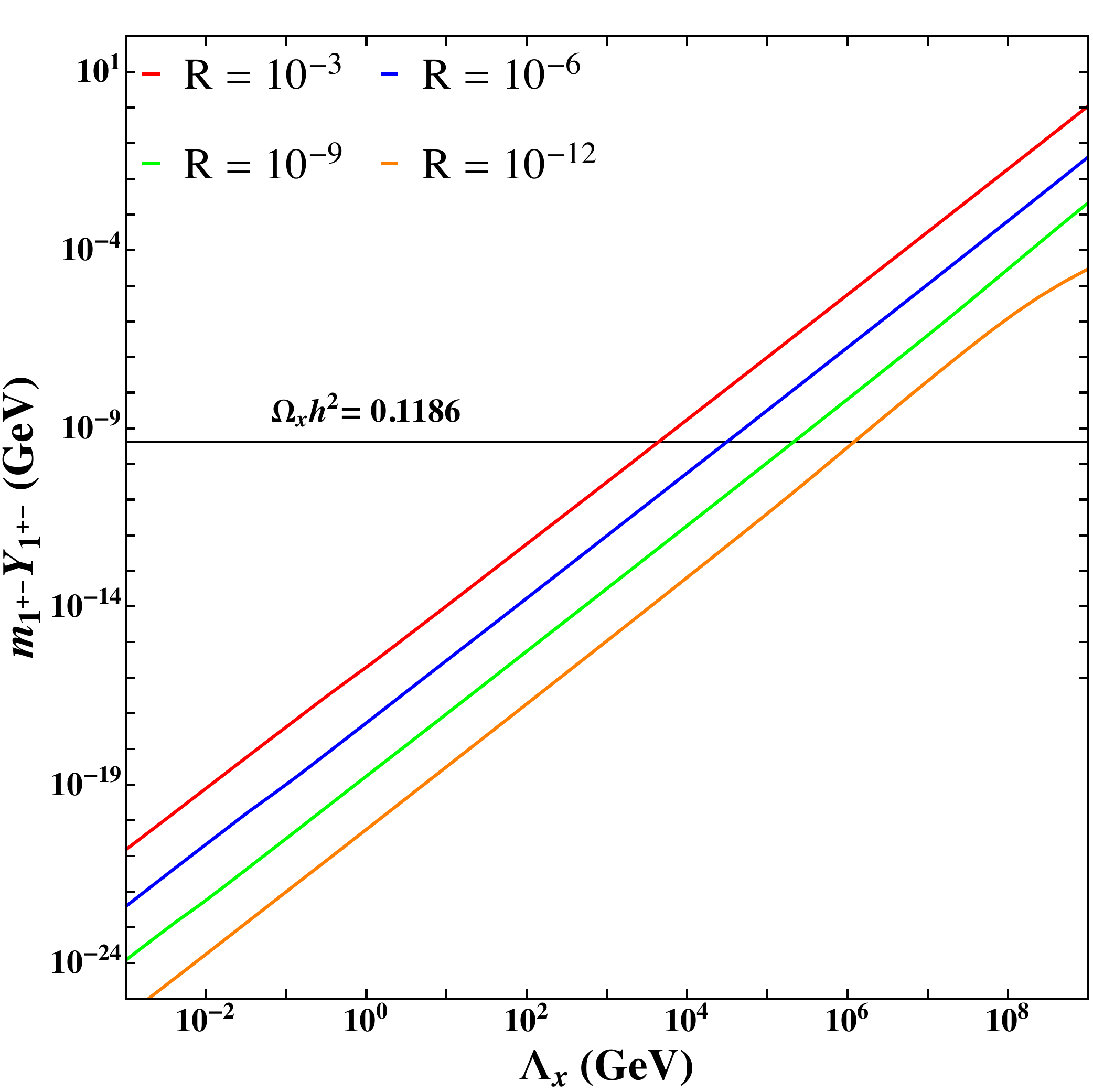}
	\hspace{0.5cm}
	\includegraphics[width=0.47\textwidth]{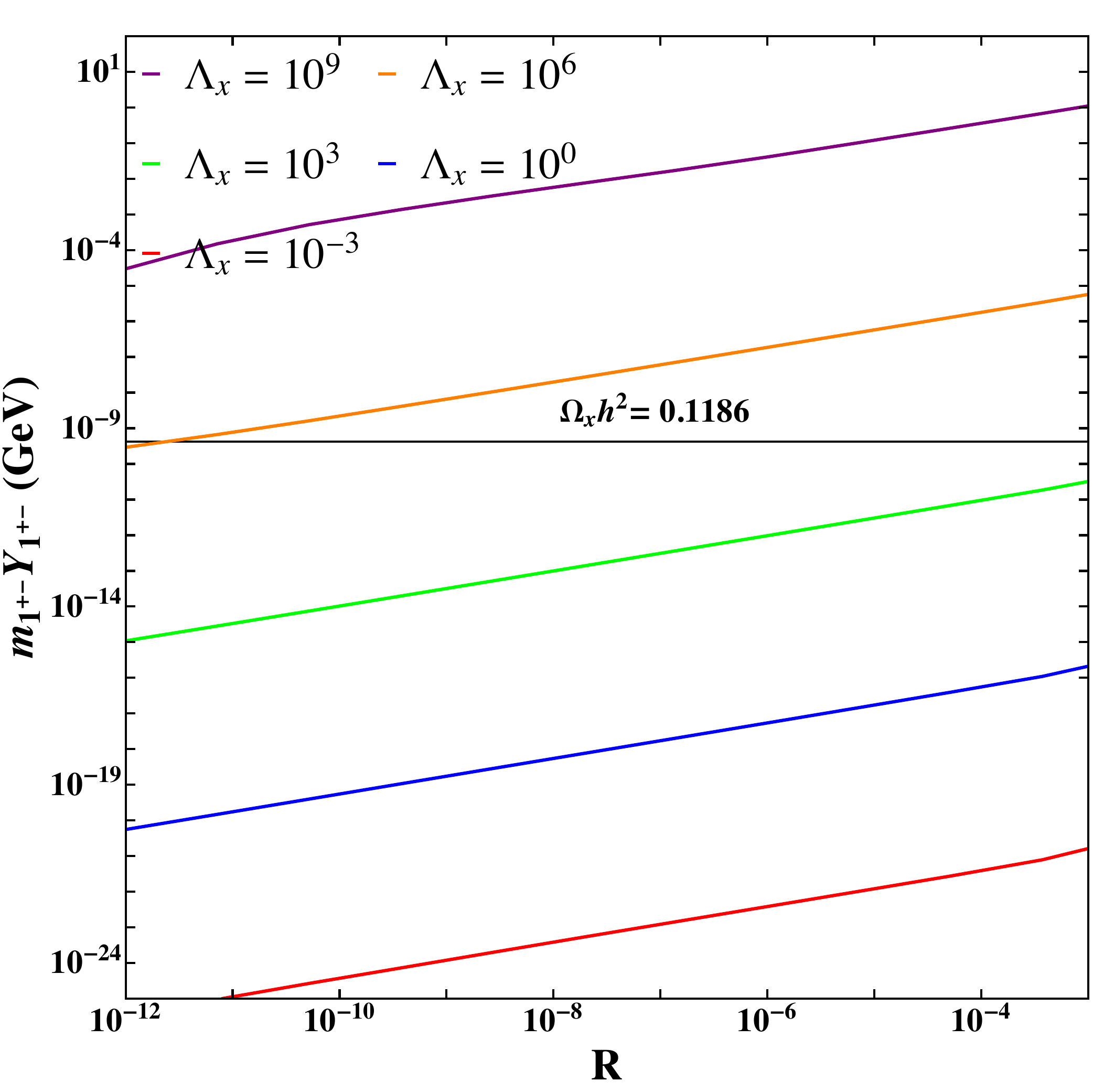}
	\hspace{0.5cm}
	\caption{
Mass-weighted relic yields of the $1^{+-}$ dark glueball in $SU(3)$ as a function of $\Lambda_x = m_x$ and $R$, computed in the simplified two-state network discussed in the text.  For reference, we also indicate the yield corresponding to the observed dark matter density.  Note that the yield of the $0^{++}$ state is much larger.
		\label{fig:oddgb}}
\end{figure}

  The mass-weighted yields of the lightest $1^{+-}$ $C$-odd $SU(3)$ glueball 
based on this analysis are shown in Fig.~\ref{fig:oddgb}.  
Like for the $C$-even states, the inclusion of additional heavier $C$-odd
glueballs generally has a negligible effect on the final abundance 
of the lightest $\zpp$ mode relative to the one-state model discussed 
previously.  Furthermore, the $\zpp$ state dominates the total 
glueball density, and the relic abundance of the $1^{+-}$ state 
is smaller by several orders of magnitude.
However, the $1^{+-}$ density can be considerably larger than any of 
the $C$-even states, even though it is heavier than the $2^{++}$ 
and $0^{-+}$ glueballs.  
As discussed above, this can be understood by the absence of relevant 
coannihilation reactions involving the much more abundant $\zpp$ glueball.
Let us also point out that $C$-odd dark glueballs provide an explicit 
realization of the scenario discussed in Ref.~\cite{Pappadopulo:2016pkp}
consisting of a stable dark matter state freezing out in the background
of a massive bath.

\section{Dark Matter Scenarios and Connections to the SM\label{sec:dm}}

  Stable dark glueballs will contribute to the dark matter~(DM) density 
of the Universe.  However, if the dark sector has a connection to the SM, 
some or all of the dark glueballs will be able to decay~\cite{Faraggi:2000pv}.  
We outline here a number of different connector scenarios and describe 
their implications for dark matter and cosmology.  A more detailed investigation
of the cosmological effects of dark glueballs will be presented 
in Ref.~\cite{us}.

\subsection{No Connection: Stable Glueballs}

With no connection to the SM, all the states in the glueball spectrum discussed
in Section~\ref{sec:gb} will be stable and contribute to the net 
DM density\footnote{Decays to gravitons are possible, but the corresponding
lifetime is much longer than the age of the Universe for 
$\Lambda_x \lesssim 10^{7}\,\gev$~\cite{Soni:2016gzf}.}.  
As reported in Sections~\ref{sec:fo} and \ref{sec:multi},
the total glueball contribution will be dominated by 
the lightest $0^{++}$ state.  The DM scenario in this case coincides
with the glueball scenarios considered inx
Refs.~\cite{Boddy:2014yra,Boddy:2014qxa,Soni:2016gzf} in which only
the lightest glueball was considered.  Avoiding overclosure by the
glueball relic density bounds $\Lambda_x$ and $R$ from above,
as can be seen in Fig.~\ref{fig:onegb}.
If the lightest glueball makes up all the DM, $\Lambda_x$ is bounded from
below by the requirement that its self-interaction cross section not be
too large, $\sigma_{2\to 2}/m \lesssim 10\,\text{cm}^2/{\text{g}}$,
which translates into~\cite{Boddy:2014yra,Boddy:2014qxa,Soni:2016gzf}
\beq 
\Lambda_x \gtrsim 100\,\mev\lrf{3}{N}^{4/3} \ .
\eeq
Smaller $\Lambda_x$ can also interfere with 
cosmic structure formation~\cite{Soni:2016gzf,deLaix:1995vi,Das:2010ts}.

\subsection{Charged Matter Connection: Unstable Glueballs}

  A minimal connection to the SM consists of heavy matter charged
under both the dark and SM gauge groups.  For example, integrating
out massive vector-like fermions at the scale $M \gg \Lambda_x$ 
generates operators of the form~\cite{Juknevich:2009ji,Juknevich:2009gg}
\beq
\lag_{eff} ~\supset~ 
\frac{\alpha_x\alpha_{i}}{M^4}\,tr(W_iW_i)\,tr(G_xG_x)
+ \frac{\alpha_x^{3/2}\alpha_Y^{1/2}}{M^4}B_{\mu\nu}\,tr(G_xG_xG_x)^{\mu\nu} \ ,
\label{eq:cmat}
\eeq
where $\alpha_x = g_x^2/(4\pi)^2$ refers to the dark gauge coupling
evaluated at the perturbative matching scale $M\gg \Lambda_x$ 
and $\alpha_i$ ($i=Y,2,3$) to the SM gauge coupling.  
The operators written in Eq.~\eqref{eq:cmat}
are schematic, and represent a set of many different Lorentz contractions;
full expressions can be found in Refs.~\cite{Juknevich:2009ji,Juknevich:2009gg}.

  The operators of Eq.~\eqref{eq:cmat} connect the dark glueballs to the
SM vector bosons after glueball confinement at $\Lambda_x$.  Together,
they allow all the would-be stable glueballs discussed in Sec.~\ref{sec:gb}
to decay either directly to the SM, or to a lighter glueball state 
and a set of SM particles~\cite{Juknevich:2009ji}.
The parametric dependences of the decay rates are
\beq
\Gamma ~\sim~ \left\{\alpha_x^2\alpha_i^2,\;\alpha_x^3\alpha_Y\right\}
N^2\frac{\Lambda_x^9}{M^8} \ ,
\label{eq:gam8}
\eeq
with significant additional suppression possible if the final-state 
phase space is constrained~\cite{Juknevich:2009ji}. 

Due to the large exponents in Eq.~\eqref{eq:gam8}, the decay rates
of glueballs through the operators of Eq.~\eqref{eq:cmat} can span 
an enormous range.  For lifetimes beyond the age of the Universe
the glueballs will contribute to the DM density and the considerations
discussed above apply here as well. In addition,
for lifetimes $\tau \lesssim 10^{26}\,\text{s}$ there will also be constraints
from energy injection into the CMB near 
recombination~\cite{Chen:2003gz,Fradette:2014sza,Yang:2015cva},
x-ray and gamma-ray fluxes~\cite{Soni:2016gzf,Cirelli:2012ut},
and energy release during primordial nucleosynthesis~\cite{Fradette:2014sza,Kawasaki:2004qu,Jedamzik:2004er,Jedamzik:2006xz}.  
Given the parametrically similar decay rates and the much larger relic
density of the lightest $\zpp$ glueball relative to the others,
these bounds apply primarily to this state.

  The operators of Eq.~\eqref{eq:gam8} can also be relevant for the glueball 
freezeout abundances.  At high temperatures they can lead
to the thermalization of the dark and visible sectors, although the
specific details depend on the reheating history after primordial inflation.  
They may also help to further populate the dark 
sector through inverse decays~\cite{Fradette:2014sza}, or induce decays 
before freezeout occurs, although this typically requires relatively larger 
values of $\Lambda_x/M$.  

  Let us also mention that heavy matter charged under both the
SM and dark gauge groups will often have accidental flavor symmetries
that lead to one or more of the new states being stable or very 
long-lived~\cite{Hill:2002ap,Kilic:2009mi,Bai:2010qg,Appelquist:2014jch,Antipin:2015xia}.  This could provide an additional contribution to the 
total density of dark matter or lead to dangerous charged relics.  
Whether such states are present and induce a cosmological problem
depends on the detailed model of the heavy matter and on the cosmological
history that we defer to a future study.  Any bounds on such states 
will also be in addition to the direct limits on glueballs we consider here.

\subsection{Higgs Portal Connection}

  A second type of SM connector is a scalar 
$\Phi_x$ charged only under $G_x$
with a Higgs portal coupling to the SM~\cite{Cline:2013zca},
\beq
\lag ~\supset~ M^2|\Phi_x|^2+ \lambda_x\,|\Phi_x|^2|H|^2 \ .
\eeq
Integrating out the scalar at its mass $M$, assumed to be
much larger than both $\Lambda_x$ and $\langle H\rangle = v$, 
produces an operator of the form
\beq
\lag_{eff} ~\supset~ 
\frac{\lambda_x\alpha_x}{M^2}|H|^2\,tr\left(G_{x\,\mu\nu}G_x^{\mu\nu}\right) \ .
\label{eq:smat}
\eeq
With $C$ and $P$ conservation in the dark sector, this is the only 
Lorentz structure generated at the dimension-six level.

After glueball confinement, the operator of Eq.~\eqref{eq:smat} gives
rise to a Higgs portal coupling between the SM Higgs boson and the 
$0^{++}$ dark glueball.  The implications of this mixing were studied
in detail in Ref.~\cite{Juknevich:2009gg} where it was shown
that it allows nearly all the heavier glueballs to decay radiatively
to lighter glueballs or directly to the SM.  If $\Delta m_x \sim \Lambda_x$
is the mass splitting for radiative decays or the mass of the decaying
glueball for direct decays to the SM, the parametric decay width 
for $\Delta m_x$ less than twice the Higgs boson mass is~\cite{Juknevich:2009gg}
\beq
\Gamma ~\sim~ 
N^2\lambda_x^2\alpha_x^2\,\frac{\Lambda_x^6}{v^2M^4}\Gamma_h(m_h=\Delta m_x) \ ,
~~~~~(\Delta M_x < 2m_h)
\label{eq:gam61}
\eeq
where $\Gamma_h$ is the width the SM Higgs would have if its mass
were $m_h = \Delta m_x$.  For $\Delta m_x$ greater than twice the Higgs
mass, the parametric width becomes
\beq
\Gamma ~\sim~
N^2\lambda_x^2\alpha_x^2\,\frac{\Lambda_x^5}{M^4} \ ,
~~~~~(\Delta m_x > 2m_h)
\label{eq:gam62} \ ,
\eeq
and is dominated by Higgs final states.

The only two ($SU(3)$) glueballs unable to decay through 
the operator of Eq.~\eqref{eq:smat}
are the $0^{-+}$ and $\opm$ states~\cite{Juknevich:2009gg}.
If the sole connection to the SM is the Higgs portal,
the dark sector has an independent charge conjugation symmetry
that implies that the lightest $C$-odd glueball is stable
(up to gravitational effects), which is $\opm$ for $G_x=SU(3)$.
The situation for the $0^{-+}$ state is more model dependent,
and decays with rates on the order of Eqs.~(\ref{eq:gam61},\ref{eq:gam62})
can arise if the visible Higgs sector contains two doublets
or if there is additional parity violation~\cite{Juknevich:2009gg}.

The DM picture that arises in this scenario consists of stable $\opm$
(and possibly $0^{-+}$) glueballs in a bath consisting mostly of metastable 
$\zpp$ states and is an explicit realization of the general scenario presented 
in Ref.~\cite{Pappadopulo:2016pkp}.
The $\zpp$ glueballs can still dominate the dark-sector contribution 
to the DM density if they are long-lived.  In contrast, if the $\zpp$
glueballs decay to the SM early enough, only the $\opm$ (and possibly $0^{-+}$)
will contribute to the DM density.  The subleading relic densities of
the heavier glueballs computed in Sec.~\ref{sec:fo} then become essential
to the net DM abundance.  Recall that the relic yields of the $\opm$
and $0^{-+}$ glueballs were computed in Sec.~\ref{sec:multi}, and are
shown in Figs.~\ref{fig:fourgb}~and~\ref{fig:oddgb}.  Of these two states,
the density of the $1^{+-}$ mode is the larger of the two.

\subsection{Yukawa Connection}

An intermediate scenario relative to the previous two can arise from 
vector-like fermions $\Psi$ and $\chi$
with both $G_x$ and $SM$ quantum numbers that couple to the
SM Higgs field according to~\cite{Juknevich:2009gg}
\beq
\lag ~\supset~ M\overline{\Psi}\Psi + M^\prime\overline{\chi}\chi
+ y_x\overline{\Psi}H\chi + (h.c.) \ .
\eeq
Integrating out these massive states at $M\sim M^\prime$, assumed to be
much larger than $\Lambda_x$ and $\langle H\rangle$, generates
the operators of Eq.~\eqref{eq:cmat} as before along with the
operator of Eq.~\eqref{eq:smat} with $\lambda_x\to y_x^2$.

This scenario opens a broad range of phenomenological possibilities.
Most of the glueballs, including the lightest $\zpp$, can decay
through either the dimension-six operator of Eq.~\eqref{eq:smat} or the
dimension-eight operators of Eq.~\eqref{eq:cmat}.  However, the $\opm$
(and possibly the $0^{-+}$) glueball is only able to decay at dimension eight.
When the ratio $\Lambda_x/M$ is large and $y_x^2/4\pi$ is not strongly 
suppressed relative to $\alpha_x$, the $\opm$ glueballs are 
parametrically long-lived relative to the $\zpp$ and other glueballs.
Thus, depending on the relative lifetimes the strongest constraints on this
scenario can come from the very late decays of the subleading
relic abundance of the $\opm$ (or $0^{-+}$) glueballs, whose yields are
shown in Figs.~\ref{fig:fourgb}~and~\ref{fig:oddgb}..

\section{Conclusions\label{sec:conc}}

  In this paper we have investigated the freezeout dynamics of $SU(3)$
dark glueballs in the early Universe.  Such glueballs arise from
confinement in theories with a new non-Abelian gauge force decoupled
from the SM and all charged matter significantly heavier than 
the confinement scale.  Our results expand upon previous studies of 
the cosmological history of dark glueballs in two key 
ways~\cite{Boddy:2014yra,Boddy:2014qxa,Soni:2016gzf}.  
First, we studied potential new effects on the glueball relic density
due to the confining phase transition itself.  And second, we performed 
a detailed analysis of the freezeout dynamics of the heavier glueballs in
the spectrum.  We also discussed connections to the SM as well as 
some of the implications of the heavier glueballs on dark matter, astrophysics, 
and cosmology, with a more detailed
study to appear in Ref.~\cite{us}.

  When the glueballs are unable to decay efficiently through connectors 
to the SM (or other lighter states), we find that the lightest $\zpp$ 
state dominates the total glueball relic abundance, 
and the abundance we calculate is in agreement with
previous studies that only considered the lightest 
state~\cite{Boddy:2014qxa,Soni:2016gzf}.  The relative relic densities
of heavier glueballs in the spectrum are orders of magnitude smaller, 
with the largest contributions coming from the $0^{-+}$ and $1^{+-}$ modes 
(for $SU(3)$).  Even though the abundances of these states
are much smaller than the lightest $\zpp$, they can also be
parametrically long-lived compared to the $\zpp$.  This opens the
possibility of the $\zpp$ mode decaying away early,
and the heavier modes making up the DM density today or leading to
the most stringent constraints on dark Yang-Mills theories.
A detailed study of these effects based on the results of this paper
is underway~\cite{us}.

  Our results are also be applicable to other simple non-Abelian gauge groups
with some straightforward modifications.  The lightest glueball,
which is generically expected to have $J^{PC} = \zpp$~\cite{West:1995ym},
will have the largest relic yield.  This yield can be computed reliably 
in the single-state model of Section~\ref{sec:fo}, provided
$3\to 2$ annihilation processes are active after the confining transition.
The relic yields of the heavier glueballs will depend on their specific
masses and quantum numbers, but can be computed following the general methods of 
Section~\ref{sec:multi}.  For a given confinement scale, 
their masses will be similar to those of $SU(3)$ for general $SU(N)$ groups,
while the $C$-odd states are expected to be considerably heavier for 
$SO(2N)$ groups and absent for $SU(2)$, $SO(2N+1)$, and $Sp(2N)$ groups
with a vanishing $d^{abc}$ symbol.  The different properties of the
more massive glueballs will only be relevant to cosmology when they
have lifetimes that are parametrically much longer than the lightest
$\zpp$ mode.

\section*{Acknowledgements}

We thank
Sonia Bacca, Nikita Blinov, Anthony Francis, Richard Hill, 
Jonathan Kozaczuk, Robert Lasenby, Randy Lewis, John Ng,
Maxim Pospelov, Adam Ritz, Josh Ruderman, Richard Woloshyn, and Yue Zhang 
for helpful discussions.  
This work is supported by the Natural Sciences
and Engineering Research Council of Canada~(NSERC), with D.~M. and K.~S.
supported in part by Discovery Grants and L.~F. by a CGS~M scholarships.
D.~M. thanks the Perimeter Institute for their hospitality while
this work was completed.  K.~S. gratefully acknowledges support from 
the Friends of the Institute for Advanced Study. 
TRIUMF receives federal funding via a contribution agreement 
with the National Research Council of Canada.  Research at 
Perimeter Institute is supported by the Government of Canada 
through Industry Canada and by the Province of Ontario through 
the Ministry of Economic Development and Innovation.


\end{document}